\documentclass[fleqn,usenatbib]{mnras}

\usepackage{newtxtext,newtxmath}
\usepackage{float}
\usepackage[T1]{fontenc}
\usepackage{ae,aecompl}
\usepackage[dvipsnames]{xcolor}
\usepackage{braket}

\usepackage{graphicx}	% Including figure files
\usepackage{amsmath}	% Advanced maths commands
\usepackage{amssymb}	% Extra maths symbols

\newcommand{\be}{\begin{equation}}
\newcommand{\ee}{\end{equation}}
\newcommand{\bea}{\begin{eqnarray}}
\newcommand{\eea}{\end{eqnarray}}
\newcommand{\rn}[1]{(\ref{#1})}
\newcommand{\pp}[2]{\frac{\partial{#1}}{\partial{#2}}}

\title[Dynamical tides and gravitational waves]{Gravitational waves from dynamical tides in white dwarf binaries}

\author[McNeill et al.]{
L.~O.~McNeill,$^{1}$\thanks{E-mail: lucy.mcneill@monash.edu} 
R.~A.~Mardling,$^{1}$
B.~M\"{u}ller $^{1}$
\\
$^{1}$Monash Centre for Astrophysics, School of Physics and Astronomy, Monash University, Victoria 3800, Australia\\
}

\date{Accepted XXX. Received YYY; in original form ZZZ}

\pubyear{2019}

\begin{document}
\label{firstpage}
\pagerange{\pageref{firstpage}--\pageref{lastpage}}
\maketitle

\begin{abstract}
We study the effect of tidal forcing on gravitational wave signals from tidally relaxed white dwarf pairs in the LISA, DECIGO and BBO frequency band ($0.1-100\,{\rm mHz}$). 
We show that for stars not in hydrostatic equilibrium (in their own rotating frames), tidal forcing will 
result in energy and angular momentum exchange between the orbit and the stars, thereby deforming the orbit
and producing gravitational wave power in harmonics not excited in perfectly circular synchronous binaries.
This effect is not present in the usual orbit-averaged treatment of the 
equilibrium tide, and is analogous to transit timing variations in multiplanet systems.
It should be present for all LISA white dwarf pairs since  
gravitational waves carry away angular momentum faster than tidal torques
can act to synchronize the spins, 
and when mass transfer occurs as it does for at least eight LISA verification binaries.
With the strain amplitudes of the excited harmonics depending directly on the density profiles of the stars,
gravitational wave astronomy offers the possibility of studying the internal structure of white dwarfs,
complimenting information obtained from asteroseismology of pulsating white dwarfs.
Since the vast majority of white-dwarf pairs in this frequency band are expected to be 
in the quasi-circular state, we focus here on these binaries, providing general analytic expressions for the dependence of the induced eccentricity and strain amplitudes on the stellar apsidal motion constants and their radius and mass ratios.
Tidal dissipation and gravitation wave damping will affect the results presented here and will 
be considered elsewhere.
\end{abstract}


%%%%%%%%%%%%%%%%%%%%%%%%%%%%%%%%%%%%%%%%%%%%%%%%%%

\section{Introduction}
Tens of millions of white dwarf binaries exist in the Milky Way \citep{Hils1990,Nelemans2004,Timpano2006}. These binaries are a promising source of low--frequency ($10^{-4}$--$10^{-1}$Hz) gravitational radiation \citep{Bender1998}, with some $10^4$ likely to be detected over the five-year LISA mission from the Milky Way \citep{Seto2002,Toonen2017}. There are also plans for other space-based detectors
(DECIGO, \citealp{Seto2001a,Kawamura2006}; BBO \citealp{Phinney2004}) that will provide better coverage for tighter
binaries in the decihertz range.
For orbital frequencies below $1\texttt{-}3 \, \mathrm{mHz}$, these binaries will mostly
contribute to the confusion-limited ``foreground''\footnote{The term ``background" is reserved for signals of cosmological origin.} noise due to the large number of systems \citep{Ruiter2010}; the less abundant
but louder sources at high frequencies will be detectable as individual resolved sources.

Most of these sources are in the galactic disc and bulge, and will have circularized well before they enter the LISA band. Such binaries form in the field together and undergo two common envelope phases, where angular momentum and mass loss leave them circular \citep{Iben1993}, although this canonical description is not consistent will all of the observed white dwarf binaries \citep{Woods2011}. 
Detailed number densities for the disc and bulge are 
given in the population synthesis study of 
\citet{Ruiter2010}. 

Due to the high degree of tidal circularization and 
the long time scale for orbital decay, 
the waveforms of most white dwarf binaries (including those which are resolvable) are expected
to be ``quasi-monochromatic'' (using
the terminology coined by \citealp{Takahashi2002}) meaning
with no discernible chirping
for
orbital frequencies $\mathord{\lesssim}10 \, \mathrm{mHz}$ \citep{Vecchio2004}
and practically all
power concentrated at twice the orbital frequency.
In this case, one would be left with the orbital period as the only system parameter that can be inferred for a
resolved binary. For this reason, scenarios that might
lead to more complex waveforms are of considerable interest.

Tidal distortion in an eccentric close binary results in advance of the periastron, while damping of the tidal motion tends to circularize and shrink the orbit, and align and synchronize the spins with the orbital motion. In the context of gravitational wave emission, previous
studies of tides in white dwarf binaries 
have focused, for example, on the circularization and merger timescales \citep{Willems2010}, spin-orbit-mode resonance locking which tends to accelerate
the merger process \citep{Burkart2013,Burkart2014}, resonant excitation of low-frequency $g$-modes and associated tidal heating \citep{Fuller2011}, and
the influence of rotation on the dynamical tide \citep{Fuller2014}. 

Of particular relevance to the present work is \citet{Willems2008}, who show that apsidal advance in {\it eccentric} white dwarf binaries should produce a measurable drift in phase of the gravitational wave signal, with power in additional harmonics that are not present for a binary point mass source. In particular, they propose to measure the apsidal motion rate and hence to constrain the internal structure of the stars via their apsidal motion constants. They argue that while the general relativistic contribution dominates this effect for orbital frequencies below $1\, \mathrm{mHz}$, thus allowing for a determination of the total mass of the system, above this frequency contributions from tides and stellar rotation start to dominate. Their analysis is based on the standard treatment of the effect of stellar quadrupole distortion from spin and tides on the orbital motion \citep[eg.][]{Sterne1939}, and therefore does not include the backreaction on the orbit. 
However, a fluid body moving in a varying gravitational field will oscillate (i.e., it will not be in hydrostatic equilibrium in its rotating frame) and that oscillation will act back on the orbit via energy and angular momentum exchange, periodically varying its eccentricity and semimajor axis and adding an oscillatory component to the apsidal motion. This will occur even after the system has reached a quasi-minimum energy state\footnote{In the sense that it can circularize no further given the system's current total angular momentum.} as long as the stellar spins are not synchronous, and/or the stars are no longer detached.
\cite{Fuller2012} have shown that 
for white dwarf binaries in the LISA band, gravitational waves carry away angular momentum more efficiently than
than tidal friction can synchronize the spins, while mass transfer is occurring for at least eight LISA verification binaries.
The influence of tidal oscillations on the gravitational wave signal may therefore be detectable in these systems.

To study this effect, we use 
the concept of {\it osculating orbital elements},\footnote{The instantaneous tangential or ``kissing'' Keplerian
orbit.}  whose instantaneous values are given by a
coordinate transformation between the six current values of the relative position and velocity coordinates,
and the six associated orbital elements. This idea has been successfully employed in the analysis
of transit timing variations, which occur when the orbit of a transiting planet is perturbed by a companion planet
\citep[eg.][]{Agol2005}. With perturbations occurring on the sub-orbital timescale, an inversion of the 
expressions involving
the osculating elements provides estimates for the perturber planet mass and its orbital elements at epoch.
In the context of the binary-tides problem, \citet{Zahn1966} appears to be the first to point out that
non-radial stellar oscillations will produce variations in the osculating orbital elements.

We also note here that that if a tidally active stellar binary has an eccentric tertiary companion, the binary can never achieve the perfectly circular synchronous state. Rather, the system will relax to a 
quasi-fixed-point state such that the binary eccentricity is non-zero and the apsidal lines are aligned or anti-aligned \citep{Mardling2007,Mardling2010}.
This in turn provides a varying gravitational field to the binary stars which will cause them to oscillate, although in this case
the gravitational wave power in the additional harmonics will come from a combination of the dynamical tide and the non-circular orbit.

Here we study this quasi-circular non-resonant effect, deriving the dependence of the induced eccentricity (and hence the gravitational wave signal) on the binary mass ratio, the ratio of stellar radii to semimajor axis, and the apsidal motion constants of the stars.
With the eccentricity forced mainly at the orbital frequency, $\nu$,
we show that while a truly circular binary produces gravitational wave power at twice the orbital frequency, a quasi-circular binary has power at $2\nu-\nu=\nu$ and $2\nu+\nu=3\nu$, with the relative amplitude at $\nu$ and $3\nu$ compared to $2\nu$ increasing like $\nu^{10/3}$ (the signal strength itself increases like $\nu^{4/3}$), and with the $3\nu$ harmonic potentially having a significant signal-to-noise ratio for at least one semi-detached LISA verification binary at 3\,mHz (Section~\ref{CV}). It is therefore, in principle, possible to constrain the internal structure of the participating stars by comparing the amplitudes of these three harmonics, and since the vast majority of LISA-band white dwarf pairs will be in the quasi-circular state, the chance that that some will be resolvable is non-negligible. In contrast, the chance of observing a recently formed tidally interacting eccentric white dwarf pair in a Milky Way globular cluster is vanishing small given the formation-rarity of such objects and their short circularization timescale (less than a million years from gravitational wave emission alone).

As we demonstrate in Section ~\ref{CV}, this effect may be evident in at least one of the verification binaries (HM Cnc at 3\,mHz),
with the signal-to-noise of the tidally forced $3\nu$ harmonic being up to 20 at a distance of 1\,kpc,
although this estimate is likely to be less when spin evolution, tidal dissipation and gravitation wave damping 
are taken into account.
There are currently 10 such verification binaries, but as noted by \citet{Kupfer2018}, this sample is observationally strongly biased and incomplete. Thus it is hoped that many other sources with similar frequencies will be discovered during the wait for LISA's launch.

Even shorter-period systems at the late stage of inspiral will be rare.
There is, however, a significant probability of catching white dwarf binaries in this state.
If type~Ia supernovae come predominantly from
the double-degenerate channel as has recently been argued \citep{Maoz2012},
then one expects to find at least
\begin{equation}
\partial N/\partial \ln f \approx R_\mathrm{Ia} \tau_\mathrm{GW}
\end{equation}
white dwarf binaries per unit orbital frequency $f$ in the Milky Way at any given
time up to several $10 \, \mathrm{mHz}$ when Roche-lobe overflow
starts. Here $R_\mathrm{Ia}$ and $\tau_\mathrm{GW}$ are the Galactic
type~Ia supernova rate and the time scale for inspiral due
to gravitational wave emission. With reasonable numbers, one obtains
$\partial N/\partial \ln f \gtrsim 0.1$, i.e.\ there is a significant
probability that LISA, DECIGO, or BBO will actually observe a double-degenerate
progenitor system.

 The paper is structured as follows: Section 2 reviews the formalism of \citet{GM1980} for the self-consistent treatment of the interaction between a binary orbit and the dynamical tide of a non-rotating star, and uses this to show that as long as the dynamical tide is active (that is, as long as the star not in hydrodynamic equilibrium), 
 the relaxed state of a close binary is quasi-circular, with expressions provided for the induced eccentricity. Section 3 considers the impact of the induced eccentricity on the gravitational wave signal including signal-to-noise and chirp, and provides examples in the form of two generic systems as well as the LISA verification source HM Cnc (RX J0806.3+1527). Finally, Section 4 presents a summary and discussion.
Notation used in the paper is listed in Table~\ref{notation}.

\section{Self-consistent treatment}

In order to determine the effect of tidal distortion on an otherwise circular orbit, we use a self-consistent treatment
which tracks the energy and angular momentum exchanged between the orbit and the (non-rotating) fluid stars.
One can then use methods of celestial mechanics to determine the time-dependent evolution of the orbital elements 
and hence the gravitational wave signal. 
Our approach is both
analytical and numerical, with the former explicitly giving the functional dependence of the gravitational wave signal
on the mass ratio, the ratio of stellar radii to semimajor axis, the apsidal motion constants of the stars and the post-Newtonian correction (and hence making the generation of a library of templates efficient), and with the latter serving to verify our results.

Our formulation is based on
the normal mode analysis set out in \cite{GM1980}, and subsequently used in \citet{RM1995a} to show
that the orbit-tide interaction can be chaotic if the orbit is eccentric and the stars are sufficiently close at periastron,
a situation which arises following tidal capture \citep{RM1995b}.
The equations used here are adapted from \citet{RM1995a}, except that we use natural units instead of the ``Chandrasekhar units''.\footnote{Chandrasekhar units are those used in the derivation of the Lane-Emden equation.}

\subsection{Equations of motion}
For the analysis we assume that only one star has finite size, while the results are easily extended to two finite-sized stars as is done later in this Section.
Moreover, since we are interested in effects associated with apsidal motion, we include the post-Newtonian correction to the equations of (relative) motion, derived from a Lagrangian and presented in \citet{kidder1995}. 
The coupled equations governing the position ${\bf r}$ of the point mass (star 2) relative to the fluid star (star 1) and the (complex) mode amplitudes
$b_{k\!l\!m}\equiv b_{\bf k}$, where $k$ is a radial mode number and $l$ and $m$ are the spherical harmonic degree and order
respectively, are
\be
\mu\ddot{\bf r}+\frac{Gm_1m_2}{r^2}\hat{\bf r}=\frac{\partial {\cal R}_{\rm tide}}{\partial{\bf r}}+\mu{\bf a}_{\rm PN}
\label{rdd}
\ee
and
\be
m_1R_1^2I_{k\!l}\left(\ddot b_{{\bf k}}+\omega_{k\!l}^2 b_{{\bf k}}\right)=\frac{\partial {\cal R}_{\rm tide}}{\partial{b_{{\bf k}}^*}},
\label{bk}
\ee
where $m_1$ and $R_1$ are the mass and radius of the fluid star, $m_2$ is the mass of the companion, 
$\mu$ is the reduced mass, $\omega_{k\!l}$ is a mode frequency,  $I_{k\!l}$ is a dimensionless structure constant
characterizing the moment of inertia of the associated tidal component (also
associated with the orthogonality properties of the eigenfunctions), and  $*$ denotes the complex conjugate.
The function
\be
\begin{split}
{\cal R}_{\rm tide}
&=
Gm_2\sum_{\bf k}c_{l\!m}\left(m_1R_1^lT_{k\!l}\right) \,b_{\bf k}\,\frac{{\rm e}^{im\psi}}{r^{l+1}}\\
&=
\frac{Gm_1m_2}{R_1}\sum_{\bf k}c_{l\!m}\,T_{k\!l} \,b_{\bf k}^*\,\frac{{\rm e}^{-im\psi}}{(r/R_1)^{l+1}},
\end{split}
\label{Rtide}
\ee
is the orbit-tide interaction energy such that the total energy
\be
E_{\rm tot}=\frac{1}{2}\mu\dot{\bf r}\cdot\dot{\bf r}-\frac{Gm_1m_2}{r}
+m_1R_1^2\sum_{\bf k}I_{k\!l}\left(\dot b_{{\bf k}}\dot b_{{\bf k}}^*+\omega_{k\!l}^2 b_{{\bf k}}b_{{\bf k}}^*\right)
-{\cal R}_{\rm tide}+E_{\rm PN}
\label{Etot}
\ee
and total angular momentum
\be
{\bf J}_{\rm tot}=\mu{\bf r}\times\dot{\bf r}-m_1R_1^2\sum_{\bf k}im\,I_{k\!l}\,b_{\bf k}\dot b_{\bf k}^*+{\bf J}_{\rm PN}
\label{Jtot}
\ee
are conserved, with
$\sum_{\bf k}=\sum_{k=1}^\infty\sum_{l=2}^\infty\sum_{m=-l,2}^l $ ($m$ is in steps of 2).
Here
\be
\begin{split}
& E_{\rm PN}
=\frac{1}{2}\mu a^2\nu^2\left(\frac{a\nu}{c}\right)^2 \times \\ &\left[
\frac{3}{4}(1-3\eta)\left(\frac{\dot{\bf r}\cdot\dot{\bf r}}{a^2\nu^2}\right)^2
+(3+\eta)\left(\frac{\dot{\bf r}\cdot\dot{\bf r}}{a^2\nu^2}\right)\left(\frac{a}{r}\right)
+\eta\left(\frac{\dot r}{a\nu}\right)^2\left(\frac{a}{r}\right)
+\left(\frac{a}{r}\right)^2
\right]
\end{split}
\label{RPN}
\ee
and
\be
{\bf J}_{\rm PN}=(\mu{\bf r}\times\dot{\bf r})\left(\frac{a\nu}{c}\right)^2\left(\frac{1}{2}(1-3\eta)\left(\frac{\dot{\bf r}\cdot\dot{\bf r}}{a^2\nu^2}\right)^2+(3+\eta)\left(\frac{a}{r}\right)\right),
\ee
with $a$ the semimajor axis, $\nu$ the orbital frequency, $\eta=m_1m_2/(m_1+m_2)^2$,
$r=|{\bf r}|$, $\psi=f+\varpi$ the true longitude with $f$ the true anomaly and $\varpi$ the longitude of periastron,

$c_{l\!m}=[4\pi/(2l+1)]Y_{l\!m}(\pi/2,0)$ with $Y_{l\!m}(\theta,\psi)$ a spherical harmonic, quadrupole values being $c_{20}=-\sqrt{\pi/5}$
and $c_{22}=c_{2-2}=\sqrt{3\pi/10}$, and
the dimensionless structure constant $T_{k\!l}$ is associated with mass-moment integrals over the fluid star.
Note that $b_{kl-m}=(-1)^mb_{k\!l\!m}^*$, $c_{l-m}=(-1)^mc_{l\!m}$, and the gradient appearing in \rn{rdd}
is such that $\partial/\partial{\bf r}={\bf e}_r\partial/\partial{r}+({\bf e}_\psi/r)\partial/\partial{\psi}$
with ${\bf e}_r$ and ${\bf e}_\psi$ plane polar unit vectors. The post-Newtonian perturbing acceleration appearing in \rn{rdd} as given by \citep{kidder1995} is
\be
\begin{split}
{\bf a}_{\rm PN}
&=-a\nu^2\left(\frac{a\nu}{c}\right)^2\left(\frac{a}{r}\right)^2\\
&\times\left\{\left[(1+3\eta)\left(\frac{\dot{\bf r}\cdot\dot{\bf r}}{a^2\nu^2}\right)-2(2+\eta)\left(\frac{a}{r}\right)-\frac{3}{2}\eta\left(\frac{\dot r}{a\nu}\right)^2
\right]{\bf e}_r\right.\\
&\left.\hspace{4cm}-(2-\eta)\left(\frac{\dot r\dot{\bf r}}{a^2\nu^2}\right)
\right\}.
\end{split}
\ee
Note that $\partial E_{\rm PN}/\partial{\bf r}\ne -\mu{\bf a}_{\rm PN}$ because the canonical momentum associated with the Lagrangian used to derive ${\bf a}_{\rm PN}$ is not equal to $\mu\dot{\bf r}$.

One can show that $T_{k\!l}$ and $I_{k\!l}$ are related to the structure constants $Q_{k\!l}$ of \citet{Lee1986} (and subsequent
studies using a similar formalism) by $Q_{k\!l}^2=T_{k\!l}^2/I_{k\!l}$, a combination which appears in the analysis below.
In particular, we show that $T_{12}^2/(\overline\omega_{12}^2I_{12})$ is proportional to the apsidal motion constant $k_2^{(1)}$,
and as such, include only
the dominant quadrupole ($l=2$) $f$-mode ($k=1$) in this study.
The structure constants
$I_{12}$, $T_{12}$ and $\overline{\omega}_{12}^2\equiv \omega_{12}^2/(Gm_1/R_1^3)$ are listed in Table~\ref{tab:modes}

\begin{table}
\caption{Structure constants}
\label{tab:modes}
\begin{tabular}{ccccc}
$I_{12}$ & $T_{12}$ & $\overline\omega_{12}^2$ & $Q_{12}$ & $(T_{12}^2/I_{12})/Q_{12}^2$\\ \hline
$6.145\times 10^{-3}$ & $ 3.848\times 10^{-2}$ & 2.120  &  0.4909 &  1.000 \\
\end{tabular}
\label{tab:constants}
\end{table}
for an $n=1.5$ polytrope.
Also listed are $Q_{12}$ and $(T_{12}^2/I_{12})/Q_{12}^2$, the latter being close to unity as it should be.

\subsection{Eccentricity and apsidal variation for quasi-circular orbits}\label{section:eccw}

As long as the stars in a white dwarf binary are not in hydrostatic equilibrium in their individual rotating frames,
they will oscillate. Thus at best the orbit will be quasi-circular, producing 
gravitational wave power in harmonics not present in perfectly circular binaries.
This must also be the case when Roche-lobe overflow occurs, especially if accretion is directly onto the star
(that is, there is no room for an accretion disk) as is
thought to be the case for the LISA verification binary HM Cnc \citep{Barros2007}; as long as a star does not fill
a closed equipotential surface, it cannot be in hydrostatic equilibrium and hence must oscillate in response
\citep{Eggleton2011}. Note that at least eight LISA verification binaries are in the semi-detached state \citep{Kupfer2018}.
Exchange of energy and angular momentum between the oscillating stars and the orbit will produce variations
in the osculating orbital elements, in particular, the eccentricity and corresponding apsidal angle which are
primarily associated with angular momentum exchange, and the semimajor axis which is primarily associated with energy exchange.
Since the latter is of order (induced) eccentricity squared,\footnote{From \rn{Etot} and \rn{Jtot}, the oscillation energy
and angular momentum are such that $E_{\rm osc}\propto J_{\rm osc}$. Moreover, $\delta E_{\rm tot}/E_{\rm tot}=-\delta a/a$
and $\delta J_{\rm tot}/J_{\rm tot}=\delta a/2a-e\delta e/(1-e^2)$ and the result follows.}
we ignore the effect on the semimajor axis and focus on variations in the eccentricity and apsidal angle.
The following analysis allows us to determine the functional dependence of these elements 
on the system parameters, which will then be used in Section~\ref{section:GWsignals} to calculate the gravitational wave amplitudes.
Note that for now we retain summation over the modes so that the notation remains compact.

Since the eccentricity and apsidal orientation are directly coupled, a consistent way to track their variation 
is to define the {\it complex eccentricity} \citep[eg.][]{Laskar12}
\be
z=e\,{\rm e}^{i\varpi},
\ee
which is equivalent to the Runge-Lenz vector \citep[eg.][]{goldstein}
but which has the advantage of being considerably simpler to work with.
The variation of the eccentricity and orientation of the binary can then be found as follows.

First note that the following quantity appearing in \rn{Rtide}
can be expressed as a Fourier series in the orbital frequency such that
\be
\begin{split}
\frac{{\rm e}^{im\psi}}{(r/a)^{l+1}}
&=
{\rm e}^{im\varpi}\frac{{\rm e}^{imf}}{(r/a)^{l+1}}\\
&=
{\rm e}^{im\varpi}\sum_{n=-\infty}^\infty X_n^{-(l+1),m}(e)\,{\rm e}^{inM}\\
&=
\sum_{n=-\infty}^\infty X_n^{-(l+1),m}(e)\,{\rm e}^{i(m-n)\varpi}\,{\rm e}^{in\lambda},
\end{split}
\ee
where $M$ is the mean anomaly, $\lambda=M+\varpi=\nu t+\lambda(0)$ is the mean longitude, and
\be
\begin{split}
X_n^{-(l+1),m}(e)
&=
\frac{1}{2\pi}\int_0^{2\pi} \frac{{\rm e}^{-imf}}{(r/a)^{l+1}}\,{\rm e}^{inM}\,dM\\
&\equiv
x_n^{-(l+1),m}\,e^{|m-n|}+{\cal O}\left(e^{|m-n|+2}\right)
\label{hansen}
\end{split}
\ee
is a {\it Hansen coefficient} \citep[eg.][]{murray}, with $x_n^{-(l+1),m}=x_{-n}^{-(l+1),-m}$ the coefficient of the leading non-zero term
in a power series expansion of $X_n^{-(l+1),m}(e)$ \citep{Mardling2013}. The values of $x_n^{-(l+1),m}$
relevant here are listed in Table~\ref{xn}.

\begin{table}
\caption{Values of $x_n^{-(l+1),m}$ for $l=2$ and $m=0,2$.}
\label{xn}
\begin{tabular}{ccccccc}
$n$ & $-1$ & 0 & 1 & 2 & 3 & 4\\ \hline
&&&&&\\
$x_n^{-3,0}$ & 3/2 & 1 & 3/2 & $-1/4$ & & \\
$x_n^{-3,2}$ &  & 5/2 & $-1/2$ & 1 & 7/2 & 1 \\
\end{tabular}
\end{table}
To first order in eccentricity and hence $z$ and $z^*$, the orbit-tide interaction energy \rn{Rtide} then becomes
\be
\begin{split}
{\cal R}_{\rm tide}=
\mu a^2\nu^2 \sum_{\bf k}c_{l\!m}\,T_{k\!l} \,b_{\bf k}\left(\frac{R_1}{a}\right)^l
&\left[x_{m-1}^{-(l+1),m}\,z\,{\rm e}^{i(m-1)\lambda}\right.\\
+{\rm e}^{im\lambda}
&\left.+x_{m+1}^{-(l+1),m}\, z^*\,{\rm e}^{i(m+1)\lambda}\right],
\end{split}
\label{calR2}
\ee
while the mode amplitudes are governed by
\be
\begin{split}
\ddot b_{{\bf k}}&+\omega_{k\!l}^2 b_{{\bf k}}
=
\left(\frac{Gm_1}{R_1^3}\right) \left(\frac{m_2}{m_1}\right)c_{l\!m}\left(\frac{T_{k\!l}}{I_{k\!l}}\right)\left(\frac{R_1}{a}\right)^{l+1}\\
&\times\left[x_{m-1}^{-(l+1),m} z^*\,{\rm e}^{-i(m-1)\lambda}
+{\rm e}^{-im\lambda}
+x_{m+1}^{-(l+1),m} z\,{\rm e}^{-i(m+1)\lambda}\right].
\end{split}
\label{bdd}
\ee

For small eccentricities, the second term in the square brackets in \rn{bdd} dominates forcing of the mode amplitudes,
so that neglecting the terms involving $z$ and $z^*$ gives
\be
b_{\bf k}(t)
=
C_1 {\rm e}^{i\omega_{k\!l} t}+C_2 {\rm e}^{-i\omega_{k\!l} t}
+c_{l\!m}\left(\frac{m_2}{m_1}\right)\left(\frac{T_{k\!l}}{I_{k\!l}}\right)\left(\frac{R_1}{a}\right)^{l+1}
\frac{{\rm e}^{-im\lambda}}{\overline\omega_{k\!l}^2-m^2\overline\nu^2},
\label{btw}
\ee
where $\overline\nu^2=\nu^2/(Gm_1/R_1^2)$,
and $C_1$ and $C_2$ are arbitrary constants.
This becomes
\be
b_{\bf k}(t)
=
c_{l\!m}\left(\frac{m_2}{m_1}\right)\left(\frac{T_{k\!l}}{I_{k\!l}}\right)\left(\frac{R_1}{a}\right)^{l+1}
\frac{{\rm e}^{-im\lambda}}{\overline\omega_{k\!l}^2-m^2\overline\nu^2}
\label{bt}
\ee
in the presence of mode damping, that is, the transient contributions associated with free oscillations die away,
leaving only forced oscillations.
To zeroth-order in $z$,
the rate of change of the complex eccentricity is then governed by (see Appendix~\ref{appendix:zdot})
\be
\label{zdot}
\begin{split}
\dot z
&=
\frac{2i}{\mu\nu a^2}\pp{{\cal R}_{\rm tide}}{z^*}+\dot z_{\rm PN}\\
&=
2i\nu\left[
\sum_{\bf k}c_{l\!m}\,T_{k\!l} \left(\frac{R_1}{a}\right)^l
b_{\bf k}\,x_{m+1}^{-(l+1),m}{\rm e}^{i(m+1)\lambda}-\frac{1}{2}\left(3-\eta\right)\left(\frac{a\nu}{c}\right)^2{\rm e}^{i\lambda}\right]\\
&=
2i\nu\left[
\sum_{\bf k}\left(\frac{m_2}{m_1}\right)\left(\frac{R_1}{a}\right)^{2l+1}\left(c_{l\!m}^2\frac{T_{k\!l}^2}{I_{k\!l}} \right)
\left(\frac{x_{m+1}^{-(l+1),m}}{\overline\omega_{k\!l}^2-m^2\overline\nu^2}\right)\right.\\
&\left.\hspace{5cm}-\frac{1}{2}\left(3-\eta\right)\left(\frac{a\nu}{c}\right)^2\right]
{\rm e}^{i\lambda},
\end{split}
\ee
where we have used the expression in \rn{bt} for $b_{\bf k}$,
so that putting $\lambda=\nu t+\lambda(0)$ gives
\be
\begin{split}
z(t)&=z(0)+\left[
\sum_{\bf k}2\left(\frac{m_2}{m_1}\right)\left(\frac{R_1}{a}\right)^{2l+1}\left(c_{l\!m}^2\frac{T_{k\!l}^2}{I_{k\!l}} \right)
\left(\frac{x_{m+1}^{-(l+1),m}}{\overline\omega_{k\!l}^2-m^2\overline\nu^2}\right)\right.\\
&\hspace{1cm}\left.-\left(3-\eta\right)\left(\frac{a\nu}{c}\right)^2\right]
\left({\rm e}^{i\lambda}-{\rm e}^{i\lambda(0)}\right)
\end{split}
\label{zt0}
\ee
where $z(0)=e(0){\rm e}^{i\varpi(0)}$.

Recalling that $x_n^{l,m}=x_{-n}^{l,-m}$ and including only the quadrupole $f$-mode for which $k=1$, $l=2$, $m=-2,0,2$,
\rn{zt0} becomes
\be
z(t)=z(0)+
\left(A_{12}-A_{\rm PN}\right){\rm e}^{i\lambda(0)}\left({\rm e}^{i\nu t}-1\right),
\label{z0}
\ee
where
\be
A_{12}=\frac{3\pi}{5}\left(\frac{m_2}{m_1}\right) \left(\frac{R_1}{a}\right)^5\frac{T_{12}^2}{I_{12}}
\left[
\left(\frac{3}{\overline\omega_{12}^2-4\overline\nu^2}\right)+\left(\frac{1}{\overline\omega_{12}^2}\right)
\right]
\label{eq:A12}
\ee
and
\be
A_{\rm PN}=\left(3-\eta\right)\left(\frac{a\nu}{c}\right)^2.
\label{eq:APN}
\ee
The
osculating eccentricity is thus given by

\be
\begin{split}
e(t)
&=\sqrt{z z^*}\\
&=\left[e(0)^2+2 e(0) A_{\rm tot}\left(\cos M-\cos M(0)\right)+4A_{\rm tot}^2\sin^2(\nu t/2)\right]^{1/2},
\label{eq:ete}
\end{split}
\ee
where $M$ is the mean anomaly and $A_{\rm tot}=A_{12}-A_{\rm PN}$. 
Note that this expression is independent of the reference direction.
This reduces to 
\be
e(t)=2|A_{\rm tot}\sin(\nu t/2)|,
\label{eq:et}
\ee
for a quasi-circular orbit.
The corresponding 
time variation of the longitude of periastron for a quasi-circular orbit is

\be
\begin{split}
\varpi
&={\rm Arg}(z)={\rm Tan}^{-1}\left(\frac{\sin\lambda-\sin\lambda(0)}{\cos\lambda-\cos\lambda(0)}\right)\\
&=[\nu t/2+\lambda(0)]({\rm mod}\,\pi)-\pi/2+\beta,
\end{split}
\label{eq:pomega}
\ee
where $\beta=0$ when $A_{\rm tot}\ge 0$ and $\beta=\pi$ otherwise.
Thus $\varpi$ librates around 0 between $-\pi/2+\epsilon$ and $\pi/2-\epsilon$ when $\beta=0$, and librates around $\pi$ between $\pi/2+\epsilon$ and $3\pi/2-\epsilon$ when $\beta=\pi$, with $\epsilon>0$ and $\epsilon\rightarrow 0$ as $e(0)\rightarrow 0$ (note that for $e(0)>0$, this libration, whose amplitude decreases as $e(0)$ increases, is superimposed on the usual positive drift component). 
Therefore the induced eccentricity and accompanying apsidal advance can be regarded as a wave of deformation of the orbit
whose frequency is equal to the orbital frequency (notwithstanding the half-angle appearing in \rn{eq:et} and \rn{eq:pomega}), and which is produced
in response to the oscillating quadrupole deformation of the fluid star.

Note that while \rn{z0} accurately predicts the osculating eccentricity via 
\rn{eq:ete} for $0\le z(0)\lessapprox 0.1$, it does not capture the secular contribution to apsidal advance
which is associated with second-order terms in eccentricity in the disturbing function \rn{Rtide}.
Retaining these produces \rn{wGR} and \rn{wtide} in Appendix~\ref{appendix:zdot},
with the associated secular rate of apsidal advance given by \rn{wsec}, consistent with standard expressions
\citep[eg.][]{RM2002}.

\begin{figure}
  \centering
    \includegraphics[width=0.5\textwidth]{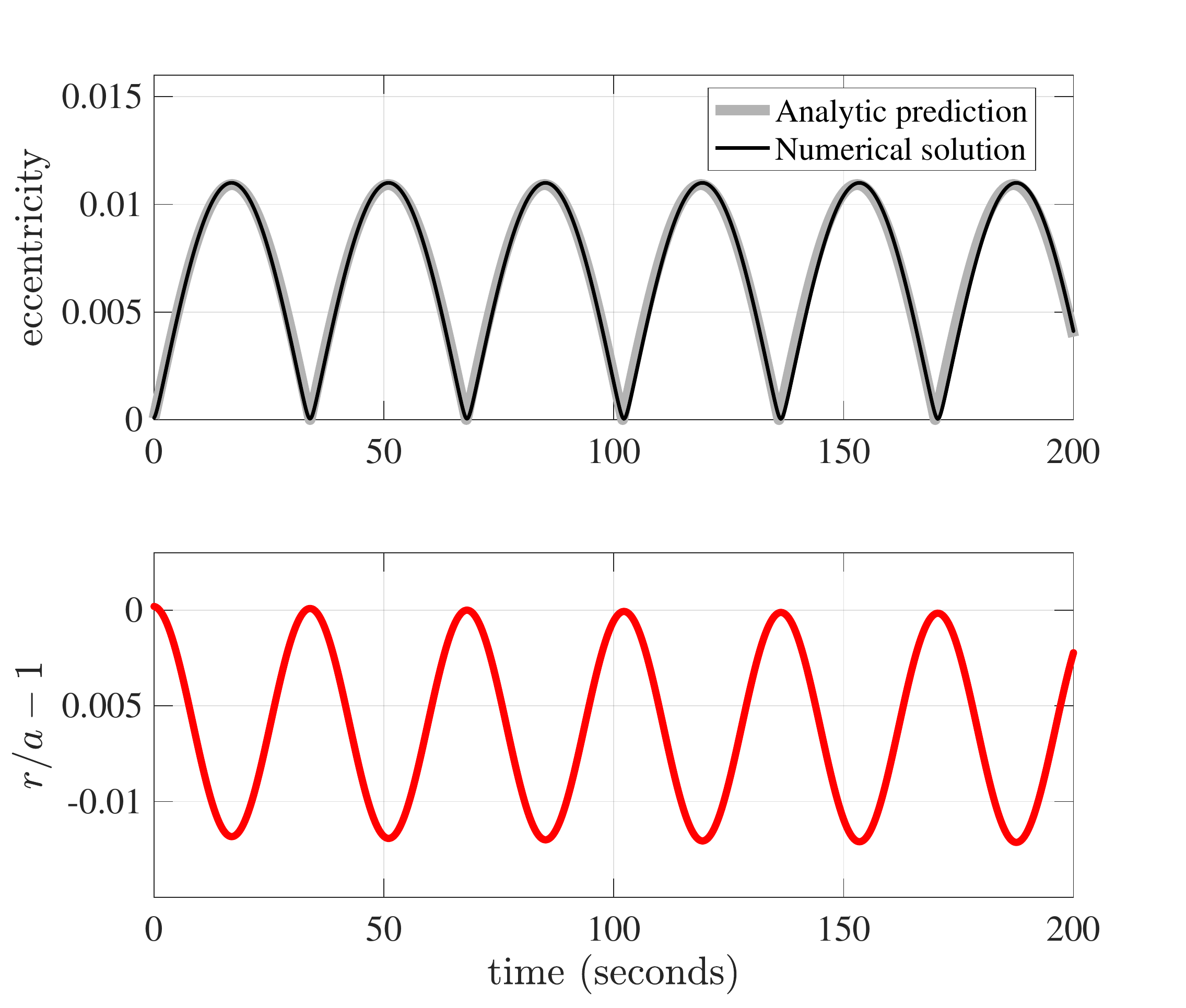}
  \caption{Evolution of the eccentricity (black) and relative change in the separation (red) over several (phase-matched) relaxed orbits for a quasi-circular pair of solar mass white dwarfs (both with radii $6\times 10^6$ m) with orbital frequency 30\,mHz (periastron separation $3.3R_1$). 
}
  \label{fig:eccev}
  \end{figure}

A comparison between the numerical and analytic solutions for the osculating eccentricity
is shown in Figure~\ref{fig:eccev} 
  for a quasi-circular pair of solar mass white dwarfs with orbital frequency 30\,mHz. The numerical solution has been artificially damped by including the term $-\dot b_{\bf k}/\tau_d$ in \rn{bdd}, where $\tau_d=\tau_d^{(0)}{\rm e}^{t/\tau_d^{(0)}}$ with $\tau_d^{(0)}= 10^4\,\mathrm{s}$. This introduces a phase lag which we have corrected for in the figure. A 4th-order Runge-Kutta integrator was used to obtain the numerical solution.
Also shown is the time dependence of the distance between the stars, given to first-order in eccentricity 
for a quasi-circular orbit by
\be
\begin{split}
r(t)&=a(1-e\cos M)\\
&=a\left[1-\frac{1}{2}\left(z^*{\rm e}^{i\lambda}+z{\rm e}^{-i\lambda}\right)\right]\\
&=a\left[1-2A_{\rm tot}\sin^2(\nu t/2)\right]
\label{rt}
\end{split}
\ee
Thus the orbit is truly non-circular as long as the dynamical tide operates. The amplitude of the effect will be modified
by tidal friction and gravitational wave emission and will tend towards a perfectly circular orbit as
the oscillations cease (for example, when the spins are perfectly synchronous in the absence
of accretion); this will be addressed in a companion paper.

\subsection{The apsidal motion constant}

The quadrupole apsidal motion constant of an object, $k_2^{(1)}$, is defined to be such that the 
quadrupole contribution to the perturbing potential due to its non-sphericity (itself induced by a companion body of mass $m_2$
and distance $a$, and/or spin distortion) is \citep{Sterne1939}
\be
\Phi_{\rm quad}=\frac{Gm_2}{R_1}\left(\frac{R_1}{a}\right)^6(2k_2^{(1)}).
\label{stern}
\ee
Substituting \rn{bt} into \rn{calR2}, retaining zeroth-order terms in eccentricity and quadrupole terms only and
taking the system to be far from resonance (ie, ignoring the contribution $4\overline\nu^2$ to the denominator), \rn{calR2} 
divided by $m_1$ becomes
\be
\frac{{\cal R}_{\rm tide}}{m_1}\simeq\frac{Gm_2}{R_1}\left(\frac{R_1}{a}\right)^6\left(\frac{4\pi}{5}\frac{T_{12}^2}{I_{k\!l}\overline\omega_{12}^2}\right).
\ee
Comparing this to \rn{stern} gives
\be
k_2^{(1)}=\frac{2\pi}{5}\frac{T_{12}^2}{I_{12}\overline\omega_{12}^2}=\frac{2\pi}{5}\left(\frac{Q_{12}}{\overline\omega_{12}}\right)^2,
\label{eq:k2}
\ee
so that from Table~\ref{tab:modes} for an $n=1.5$ polytrope we obtain $k_2^{(1)}=0.143$, which compares favourably to Stern's value of 0.145.
We also have that for systems far from resonance, the variation of the eccentricity due to tidal distortion is such that
\be
A_{12}\simeq 6k_2^{(1)}\left(\frac{m_2}{m_1}\right)\left(\frac{R_1}{a}\right)^5.
\label{A12k}
\ee

\begin{figure}
  \centering
  \includegraphics[width=0.48\textwidth]{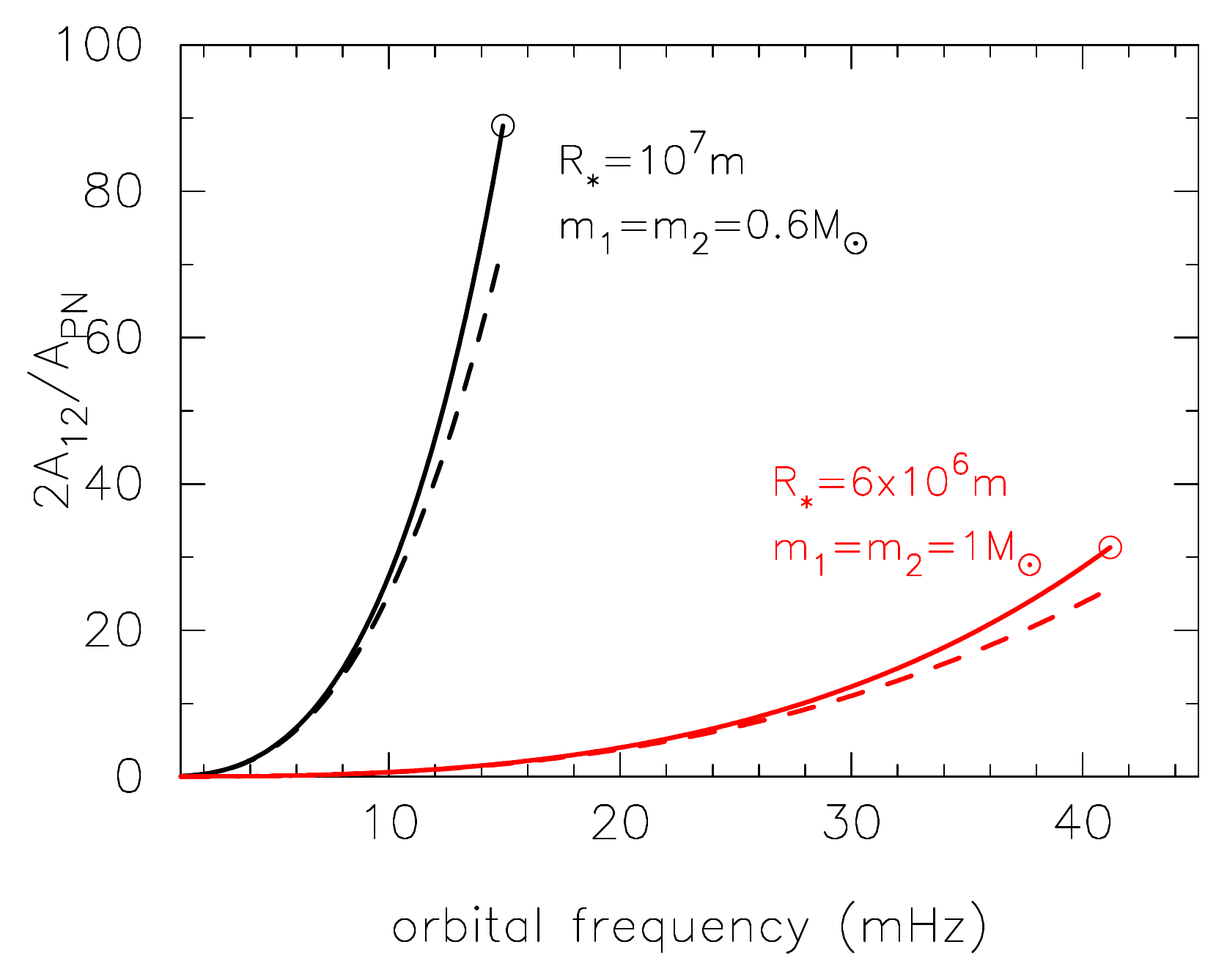}
  \caption{Relative strength of contribution to eccentricity amplitude from tides and relativity as a function of orbital frequency for two different white dwarf pairs. The dashed curves use the approximation \rn{A12k} for $A_{12}$ and demonstrate that a measurement of $A_{\rm tot}$ usefully constrains the apsidal motion constant $k_2^{(1)}$ for detectable binaries. The circles at the end of the curves indicate Roche contact.
  \label{fig:A12-APN}}
  \end{figure}
Figure~\ref{fig:A12-APN} shows the relative strength of the contributions to the eccentricity amplitude from tides and relativity as a function of orbital frequency for white dwarf pairs with $m_1=m_2=0.6 M_\odot$, $R_1=R_2=10^9\,{\rm cm}$ (black curves) and $m_1=m_2=1.0 M_\odot$, $R_1=R_2=6\times 10^8\,{\rm cm}$ (red curves). The dashed curves use the approximation \rn{A12k} for $A_{12}$ and demonstrate that a measurement of $A_{\rm tot}$ usefully constrains internal structure information, especially if the chirp mass can be measured (Section~\ref{chirp}) and electromagnetic follow-up observations can be done.  In practice, a measurement of the ratio of the strains at the $n=3$ (and/or the $n=1$) and $n=2$ harmonics constrains the quantity 
\be
6\left[k_2^{(1)}\left(\frac{m_2}{m_1}\right)\left(\frac{R_1}{a}\right)^5+k_2^{(2)}\left(\frac{m_1}{m_2}\right)\left(\frac{R_2}{a}\right)^5\right]-\left(3-\eta\right)\left(\frac{a\nu}{c}\right)^2,
\label{eq:both}
\ee
where $R_1$, $R_2$ and $k_2^{(1)}$, $k_2^{(2)}$ are the radii and apsidal motion constants of stars 1 and 2 respectively.

\subsection{Dissipation and chirp}\label{chirp}

Resolvable quasi-circular
white dwarf binaries in the LISA/DECIGO/BBO frequency band will have orbital periods in the range $3-50\,{\rm mHz}$,
with the upper bound corresponding to Roche contact of two stars at the Chandrasekhar mass limit \citep{Egg1983}.
Such an orbit will decay slightly during the observing period as a result of gravitational wave emission and tidal friction.
The associated timescales for change in the orbital frequency, $\tau_{\rm GR}$ and $\tau_{\rm tides}$, are such that
\citep{Peters1964}
\be
\tau_{\rm GR}^{-1}=\frac{96}{5}\nu\eta\left(\frac{a\nu}{c}\right)^5\left(1+\frac{157}{24}\langle e^2\rangle\right)
\label{eq:tauGR}
\ee
and \citep[equation~(55)]{RM2002}
\be
\begin{split}
\tau_{\rm tides}^{-1}
&=
\frac{171}{4}\nu\left[\left(\frac{k_2^{(1)}}{Q_1}\right)
\left(\frac{m_2}{m_1}\right)\left(\frac{R_1}{a}\right)^5
+\left(\frac{k_2^{(2)}}{Q_2}\right)
\left(\frac{m_1}{m_2}\right)\left(\frac{R_2}{a}\right)^5
\right]\langle e^2\rangle\\
&=
\frac{57}{4}\nu\left[\frac{A_{12}^{(1)}}{Q_1}+\frac{A_{12}^{(2)}}{Q_2}\right]A_{\rm tot}^2,
\end{split}
\label{eq:tautides}
\ee
where $Q_1$ and $Q_2$ are the tidal quality factors ($Q$-values) of stars 1 and 2 respectively,
and $\langle e^2\rangle=2A_{\rm tot}^2$ is the time averaged value of the square of the eccentricity
with $A_{\rm tot}=A_{12}^{(1)}+A_{12}^{(2)}-A_{\rm PN}$. Note that the form of \rn{eq:tautides} assumes
synchronous rotation of both bodies; this will be slightly modified for asynchronous rotation.
The chirp timescale, $\tau_\nu$, is then such that
\be
\tau_\nu^{-1}=\tau_{\rm GR}^{-1}+\tau_{\rm tides}^{-1},
\label{eq:tauchirp}
\ee
so that over an observing time $T$, the chirp, $\Delta f_b$, is given by
\be
\frac{\Delta f_b}{f_b}=\frac{T}{\tau_\nu},
\label{eq:Dff0}
\ee
where $f_b=\nu/2\pi$ is the orbital cyclic frequency.

Taking $Q_1=Q_2=10^7$ \citep{Burkart2013}, orbital decay is completely dominated by gravitational wave emission
for all white dwarf binaries up to Roche contact, at which point $\tau_{\rm tides}/\tau_{\rm GR}$ has reduced to 3400 and 200
for $1.0\,M_\odot$ and $0.6\,M_\odot$ pairs respectively.
Taking $\tau_\nu\simeq\tau_{\rm GR}$ one can then write
\be
\frac{\Delta f_b}{f_b}\simeq\frac{T}{100\,{\rm yr}} \left(\frac{\cal M}{M_\odot}\right)^{5/3} \left(\frac{f_b}{30\,{\rm mHz}}\right)^{8/3},
\label{eq:Dff}
\ee
where 
\be
{\cal M}=\frac{(m_1m_2)^{3/5}}{(m_1+m_2)^{1/5}}=\eta^{3/5} (m_1+m_2)
\ee
is the chirp mass,
so that a measurement of the chirp allows one to estimate ${\cal M}$ if the signal to noise is adequate.
The range of $\tau_\nu$ for a pair of solar mass white dwarfs is 58,480\,yr at 3\,mHz to 59\,yr at 40\,mHz (Roche contact),
with the corresponding chirp over 5 years of 0.003 and $3.4\,{\rm mHz}$ respectively, or double these for
the prominent $n=2$ harmonic.
For a pair of $0.6\,M_\odot$ white dwarfs the range is 137,012\,yr at 3\,mHz to 1874\,yr at 15\,mHz (Roche contact),
with corresponding chirp of $10^{-4}-0.04\,{\rm mHz}$. 

Note that white dwarf binaries close to or at contact are likely to involve some angular momentum loss from the system via mass loss, in which case the chirp will larger than predicted by \rn{eq:Dff} (see Section~\ref{CV}).

\section{Impact on the gravitational wave signal}
\label{section:GWsignals}
The dimensionless gravitational wave strain tensor is given by Einstein's quadrupole formula
\be
h_{i\!j}
=
\frac{2G}{c^4 D}\Lambda_{i\!j,k\!l}\ddot{\cal I}^{k\!l}
=
2\eta\left(\frac{a}{D}\right)\left(\frac{a\nu}{c}\right)^4\Lambda_{i\!j,k\!l}\left(\frac{\ddot{\cal I}^{k\!l}}{\mu a^2\nu^2}\right),
\ee
where ${\cal I}^{k\!l}$ is the (symmetric) moment of inertia tensor of the object emitting the gravitational waves, $D$ is its distance from the observer, and $\Lambda_{i\!j,k\!l}$ is a sky-projection operator which selects the transverse trace-less part of ${\cal I}^{k\!l}$ \citep{Maggiore2007}. Then 
\be
[h_{i\!j}]=\left[
\begin{array}{ccc}
h_+ & h_\times & 0 \\
h_\times & -h_+ & 0 \\
0 & 0 & 0
\end{array}
\right],
\label{eq:strainmatrix}
\ee
where 
\be
h_+=\frac{1}{2}h_0(\ddot {\cal I}^{11}-\ddot {\cal I}^{22})/(\mu a^2\nu^2)
\ee
and
\be
h_\times=h_0\ddot I^{12}=h_0\ddot {\cal I}^{21}/(\mu a^2\nu^2),
\ee
with $h_0 = 2\eta\left(a/D\right)\left(a\nu/c\right)^4$.
The moment of inertia tensor for a binary star system is
\be
{\cal I}^{i\!j}=\mu x^ix^j,
\ee
where $x^i$ are the components of the vector  ${\bf r}$ giving the position of $m_2$ relative to $m_1$, and $r=|{\bf r}|$. Taking the binary to be face-on to the observer, we can write
\be
{\bf r}=r(\cos\psi{\bf i}+\sin\psi{\bf j}), 
\ee
where ${\bf i}$ and ${\bf j}$ are basis vectors in the plane of the sky and $\psi$ is the true longitude measured from the ${\bf i}$ direction.
The moment of inertia tensor is then
\be
[{\cal I}^{i\!j}]=\mu r^2\left[
\begin{array}{ccc}
\cos^2\psi & \sin\psi\cos\psi & 0\\
\sin\psi\cos\psi & \sin^2\psi & 0 \\
0 & 0 & 0
\end{array}
\right]
\ee
so that
\be
h_+=h_0\left[(\dot r^2+r\ddot r-2r^2\dot\psi^2)\mathrm{cos} 2\psi-(4r\dot r\dot\psi+r^2\ddot\psi)\mathrm{sin} 2\psi\right]/(a^2\nu^2)
\label{eq:hplus}
\ee
and
\be
h_\times=
h_0\left[(\dot r^2+r\ddot r-2r^2\dot\psi^2)\mathrm{sin} 2\psi+(4r\dot r\dot\psi+r^2\ddot\psi)\mathrm{cos} 2\psi\right]/(a^2\nu^2).
\label{eq:hcross}
\ee
To first-order in the eccentricity (and hence $z$ and $z^*$) we can write \citep{murray}
\be
r/a=1-e\cos(\lambda-\varpi)
=1-\frac{1}{2}\left(z^*{\rm e}^{i\lambda}+z{\rm e}^{-i\lambda}\right),
\ee
and
\be
\psi=\lambda+2e\sin(\lambda-\varpi)
=\lambda-i\left(z^*{\rm e}^{i\lambda}-z{\rm e}^{-i\lambda}\right).
\ee
The time derivatives of $r$ and $\psi$ are obtained with $z(t)$ given by \rn{z0}, 
and \rn{eq:hplus} and \rn{eq:hcross} can be expanded to first order in $A_{\rm tot}$ to give

\be
\begin{split}
h_{+}(t) &= h_0\left[(-4+8 A_{\rm tot}^{(0)}) \cos(2\nu t+2\lambda(0))\right. \\
&\left. -3 A_{\rm tot}^{(0)} \cos(\nu t+2\lambda(0))  +9 A_{\rm tot}^{(0)} \cos(3\nu t+2\lambda(0))\right]
 \end{split}
\label{eq:hplus2}
\ee
 and     
\be
\begin{split}
h_{\times}(t) &= h_0\left[(-4+8 A_{\rm tot}^{(0)}) \sin(2\nu t+2\lambda(0))\right. \\
&\left. -3 A_{\rm tot} \sin(\nu t+2\lambda(0))  +9 A_{\rm tot}^{(0)} \sin(3\nu t+2\lambda(0))\right].
 \end{split}
\label{eq:hcross2}
\ee

\subsection{Waveform in frequency space}

In order to assess the detectability of various harmonics in the gravitational wave signal, we consider the signal in the frequency domain by calculating
the discrete Fourier transform (DFT) of the strain, which for the $k$th frequency bin is
\begin{equation}
   {h}(f_k) = \frac{1}{N} \sum_{n=1}^{N} h_{\times}(t_{n})e^{2\pi i f_k t_n}= \frac{1}{N} \sum_{n=1}^{N} h_{\times}(t_{n})e^{2\pi i k n/N},
   \label{eq:hfk}
\end{equation}
where $N$ is the number of points in the data stream and $h_\times(t_{n})$ the strain at time $t_n=nT/N$.
Note that any normalised linear combination of $h_+$ and $h_\times$ in \rn{eq:hfk} will yield the same results for the amplitude spectral density (ASD).
For a sampling frequency of $1\,\mathrm{Hz}$ and an observing time $T$ of 5~years,
the frequency resolution of the data stream, $\Delta f=1/T$, is $6\times 10^{-9}\,{\rm Hz}$.
The ASD of the (Fourier transformed) signal $h(f_k)$ is then 
\begin{equation}
\tilde{h}(f_k) = \sqrt{\frac{|{h}(f_k)|^2}{\Delta f}} = |{h}(f_k)| \sqrt{T},
\label{eq:htilde}
\end{equation}
which can be compared to the detector noise as follows.

Using an optimal matched filter, the signal-to-noise  ratio, $\mathcal{S}/\mathcal{N}$, of a continuous signal $h_\times(t)$, whose continuous Fourier transform is $H(f)$ (units Hz$^{-1}$), is given by \citep{Flanagan1998}
\begin{equation}
(\mathcal{S}/\mathcal{N})^2 = 4 \int_0^\infty \mathrm{d}f \frac{| H(f) |^2}{S(f)},
\label{eq:SNcont}
\end{equation}
where $S(f)$ is the noise power spectral density (units Hz$^{-1}$). For a discrete data stream and with our definition \rn{eq:hfk} of the DFT, the signal-to-noise ratio is instead given by \citep{Moore2015}
\begin{equation}
(\mathcal{S}/\mathcal{N})^2 = 
4 \sum_{k=0}^{N-1} \frac{|\tilde{h}(f_k) |^2}{S_k(f) }
=4 T\sum_{k=0}^{N-1} \frac{|{h}(f_k) |^2}{S_k(f)}.
\label{eq:SNdiscrete}
\end{equation}

If we disregard the shrinking of the orbit due to the radiation reaction term,
one can calculate the ASD (and hence the detectability of the individual harmonics) analytically from Equations~(\ref{eq:hplus2}) and (\ref{eq:hcross2}). For each of the harmonics, the power will be concentrated in a single frequency bin\footnote{In general, there will be a small amount of spectral leakage depending on whether or not the frequency of the harmonics coincides exactly with the discrete DFT frequencies, but this is irrelevant for the calculation of the signal-to-noise ratio.}, with an
ASD of $|4-8 A_{\rm tot}| h_0 \sqrt{T}$ for the second harmonic,
$3A_{\rm tot} h_0 \sqrt{T}$ for the first harmonic, and 
$9A_{\rm tot} h_0 \sqrt{T}$ for the third harmonic. Hence the
signal-to-noise ratio is
$\mathcal{S}/\mathcal{N}=|4-8 A_{\rm tot}| h_0 \sqrt{T/S(2f_b)}$
for the detection of the second harmonic
as an individual waveform component, and similarly for
the other two harmonics.

In reality, the decay of the orbit will introduce a chirp in the waveform and broaden the spectrum slightly. However, the signal-to-noise ratio is
not affected by this chirp (as long as the exact waveform is still available for constructing the optimal Wiener filter): in contrast with the late inspiral
phase of neutron star and black hole binaries as observed by Advanced LIGO
and VIRGO, the broadening of the individual harmonics is still small, so that for a waveform with a single harmonic at frequency $nf_b$, $n=1,2,3$, we obtain 
\begin{equation}
(\mathcal{S}/\mathcal{N})_\mathrm{chirp}^2 = 4 \int \mathrm{d}f \frac{| H_\mathrm{chirp}(f) |^2}{S(f)} \approx  \frac{4}{S(nf_b)} \int \mathrm{d}f | H_\mathrm{chirp}(f) |^2
\end{equation}
for signal-to-noise $(\mathcal{S}/\mathcal{N})_\mathrm{chirp}^2$ of the chirping waveform, since the detector noise does not vary appreciably over the frequency range
traversed by the chirping binary. Here $H_\mathrm{chirp}(f)$ is the continuous Fourier transform of \rn{eq:hcross2} with $\nu t$ replaced by $\nu_0 t+\frac{1}{2}t^2/\tau_\nu$,
where $\nu_0$ is the initial orbital frequency and $\tau_\nu$ is given in \rn{eq:tauchirp}, yielding
\be
H_{\rm chirp}(f)=
\frac{h_0}{2}\left[(-4+8A_{\rm tot}){\cal I}_2(f)-3A_{\rm tot}{\cal I}_1(f)+9A_{\rm tot}{\cal I}_3(f)\right],
\label{eq:Hchirp}
\ee
with
\be
{\cal I}_n(f)
=\left\{\begin{array}{ll}
\displaystyle{\left(\frac{\tau_\nu}{T}\right)\sqrt{\frac{1}{2nf_b\tau_\nu}}{\rm e}^{-i\phi_n}
}, & a_n(f)b_n(f)<0,\\
&\\
0, &a_n(f)b_n(f)\ge0,
\end{array}
\right.
\ee
\be
a_n(f)=-\left(f/f_b-n\right),
\hspace{0.5cm}
b_n(f)=nT/\tau_\nu-(f/f_b-n),
\label{anbn}
\ee
and 
\be
\phi_n=(\pi f_0\tau_\nu/n)(f/f_0-n)^2+\pi/4.
\ee
Here we have used the method of stationary phase to evaluate the integrals associated with the damping term. The approximation used picks up the box-like structure of the chirp, correctly giving its maximum value and width but not producing the splayed structure at low ASD.
Thus, the  signal-to-noise only
depends on the total power in the harmonic, which can also be
computed from the waveform in the time domain by dint of Parseval's
theorem,
\begin{equation}
(\mathcal{S}/\mathcal{N})_\mathrm{chirp}^2=\frac{1}{\pi S(nf_b)} \int \mathrm{d}t | h_\mathrm{chirp}(t) |^2
=\frac{T \langle | h_\mathrm{chirp}(t) |^2\rangle}{\pi S(nf_b)}.
\label{eq:SNRchirp}
\end{equation}
Since the change in orbital separation over realistic
integration periods is small, the
mean-square average $\langle | h_\mathrm{chirp}(t) |^2\rangle$
of the strain is essentially the same with and without
the radiation reaction term. The case of several discrete
harmonics is no different, and we can therefore work
directly with the amplitudes from
Equations~(\ref{eq:hplus2}) and (\ref{eq:hcross2}) to
assess the detectability of the individual harmonics.
As such, we define the ``effective ASD'' as the ASD without chirp, and use this to assess the signal-to-noise ratio
of a detection.

\begin{figure}
  \centering
  \includegraphics[width=0.5\textwidth]{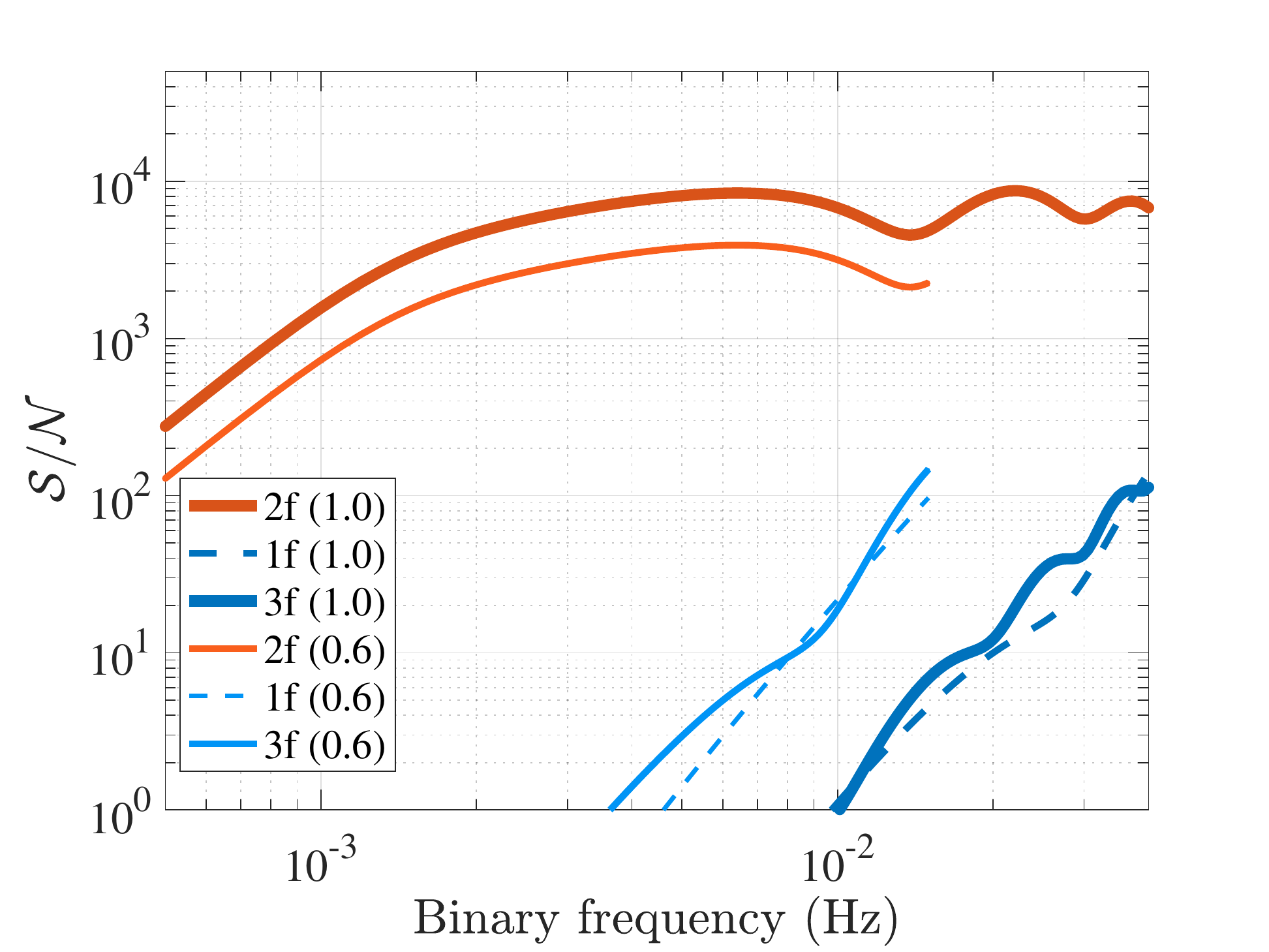}
  \caption{Signal-to-noise ratio (calculated using the ``effective ASD'') as a function of binary frequency up to Roche contact
  for second (red), third (solid blue) and first (dashed blue) orbital harmonics, for a pair of $1\,M_\odot$ white dwarfs with radii $6\times 10^8\,{\rm cm}$ (thick lines), and 
a pair of $0.6\,M_\odot$ white dwarfs with radii $10^9\,{\rm cm}$ (thin lines). A distance of 1 kpc is assumed together with
an integration time of 5\,yr. 
The tidally induced first and third harmonics rise above a LISA detection threshold of 8
at orbital frequencies 15\,mHz and 7\,mHz respectively.
}
  \label{fig:SNboth}
  \end{figure}

Figure~\ref{fig:SNboth}
shows the signal-to-noise ratio as a function of binary frequency up to Roche contact
for the first three harmonics associated with 
a pair of $1\,M_\odot$ white dwarfs with radii $6\times 10^8\,{\rm cm}$ (thick lines), and 
a pair of $0.6\,M_\odot$ white dwarfs with radii $10^9\,{\rm cm}$ (thin lines) (Sections~\ref{section:example} and \ref{section:example2}).
A distance of 1\,kpc is assumed together with
an integration time of 5\,yr.
The tidally induced first and third harmonics rise above a LISA detection threshold of 8
at orbital frequencies 15\,mHz and 7\,mHz respectively.
Both cases are plotted up to Roche contact.

\subsection{Matched filter templates}

Equation \rn{eq:Hchirp} shows that for a given observing time $T$, templates for the three-harmonic 
box-like spectra described here involve 4 free parameters:
the orbital frequency $f_b$, the amplitudes of the $n=2$ and 3 harmonics and the width of $n=2$ (or $n=3$) harmonic.
These in turn can be used to solve for $h_0$, $A_{\rm tot}$ and $\tau_\nu$.
The minimum requirement for $\tau_\nu$ to be measurable is that
$nT/\tau_\nu>\Delta f/f_b$, where again $\Delta f$ is the frequency resolution given by the sampling rate over the observing time
(see equation~\rn{anbn}),
that is, a finite chirp width must be measurable for one of the harmonics.
Thus according to equation~\rn{eq:Dff}, in the case that $\tau_\nu$ is dominated by gravitational wave emission and
for a sampling frequency of 1\,Hz and an observing baseline of 5 years, the chirp of the $n=2$ harmonic is
(in principle) measurable if the orbital frequency is greater than 0.5\,mHz.
If the chirp cannot be measured, only $A_{\rm tot}$ and the combination $h_0\sqrt{\tau_\nu}$ can be solved for.

Finally, note that while the amplitudes are affected by chirp (whether or not it can be measured),
the ratio of amplitudes is not.

\subsection{Signal from a high--mass 30\,mHz quasi--circular binary}
 \label{section:example}
 
\begin{figure}
\centering
    \includegraphics[width=0.5\textwidth]{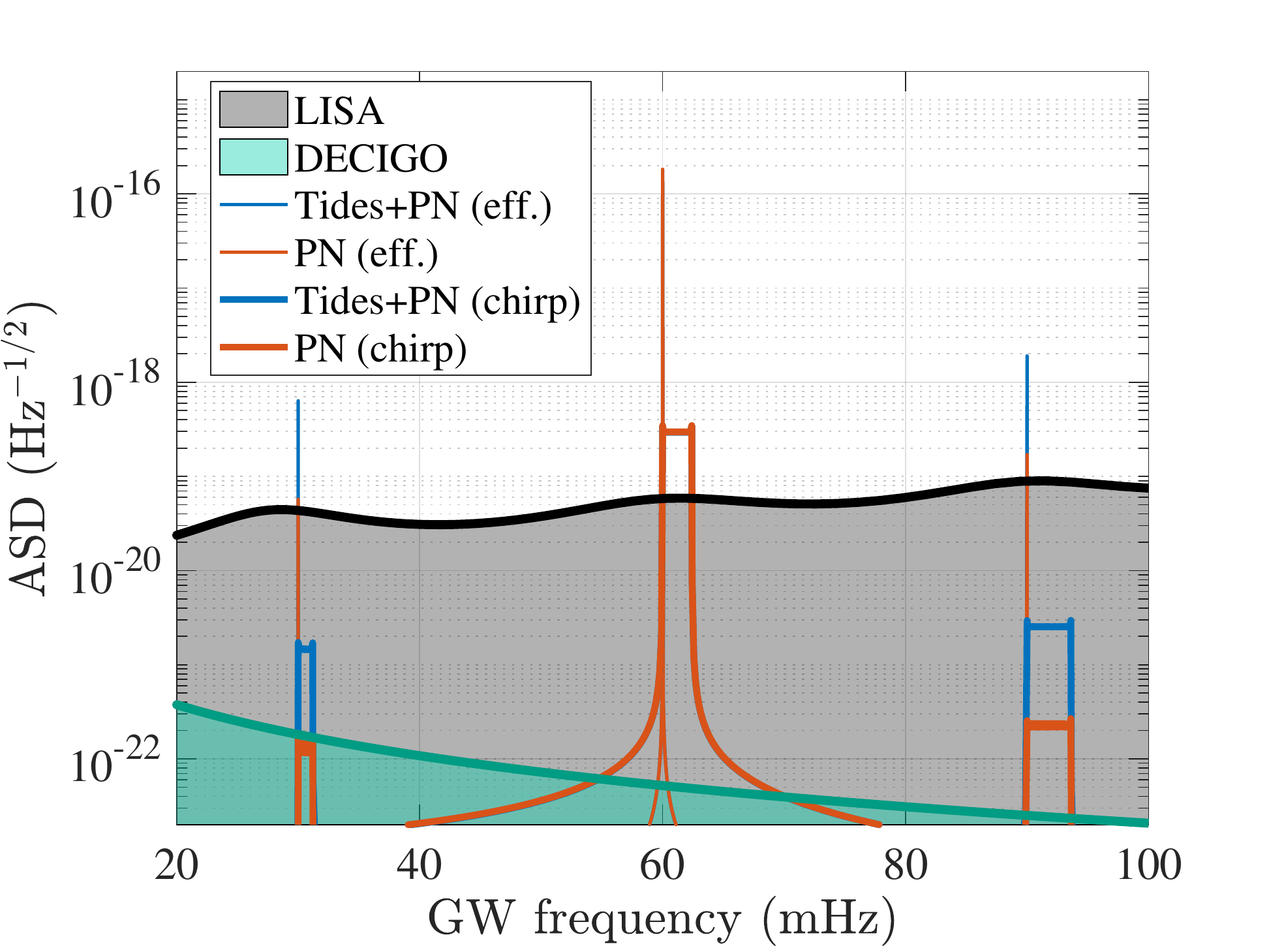}
  \caption{
  The ASD (with chirp: box-like structures) and ``effective ASD'' (no chirp: individual spikes)
  for a quasi-circular pair of $1\,M_\odot$ cold white dwarfs with orbital frequency 30\,mHz, radii $6\times 10^8\,{\rm cm}$, at a distance of 1\,kpc and for 5 years of detector integration. Red indicates the signal with PN terms only, while blue includes
both tides and GR. Since the chirp is small, the signal-to-noise of the true chirping signal (using the analytical form
\rn{eq:Hchirp} as a template) is effectively the same as
that for the signal with no chirp, allowing one to use the effective ASD to estimate this quantity. Relative to the LISA noise ASD we obtain $\mathcal{S}/\mathcal{N}$
ratios of 30, 6000 and 40 for the $n=1,2,3$ harmonics respectively, making the tidal contribution
to the signal resolvable and loud for LISA.
See text for details.}
   \label{fig:30mHz}
\end{figure}

Figure~\ref{fig:30mHz} 
shows the amplitude spectral density for a quasi-circular pair of $1\,M_\odot$ cold white dwarfs with orbital frequency 30\,mHz, radii $6\times 10^8\,{\rm cm}$ \citep{Timmes}, at a distance of 1 kpc and for 5 years of detector integration. The signal is plotted with and without chirp: the box-like structures are with chirp, while the single spikes are without and indicate
the ``effective ASD''.
Red is used for the signal without tides (but including the post-Newtonian correction), while blue includes both tides and PN.
The curved structures visible around $60\,{\rm mHz}$ are a result of using a discrete Fourier transform, here calculated 
using a Fast Fourier Transform routine; the analytic forms for both the chirped (equation~\rn{eq:Hchirp})
and non-chirped (paragraph following eqn.~\rn{eq:SNdiscrete}) signals accurately predict the amplitudes, without
the (irrelevant) curved structures.
Also shown is the LISA sensitivity ASD \citep{Larson2000} (black/grey) which is based on five years of observation, as well as 
the DECIGO noise \citep[green;][]{Yagi2011}.
Note that for this system the orbital separation is $a=3.26 R_*$, thereby avoiding Roche contact by 23\% \citep{Egg1983}, while
the orbital decay timescale due to radiation reaction is 125 yr \citep{Peters1964}, 

yielding a chirp of $2.4\,{\rm mHz}$ for
the $n=2$ harmonic over the 5 years.

Assuming an optimal matched filter, the signal-to-noise ratio of each harmonic can be estimated using the excess of effective ASD over the noise floor, multiplied 
by a factor of 2 (equation~\rn{eq:SNcont} or \rn{eq:SNdiscrete}).
For LISA we obtain ${\cal S}/{\cal N}=30$, 6000 and 40 for the $n=1,2,3$ harmonics respectively, while for DECIGO
the corresponding values are 6000, $8\times 10^6$ and $2\times10^5$.
Since no foreground strain from Galactic white dwarf pairs is expected at these frequencies \citep{Ruiter2010,Bender1998}, 
such a signal should be resolvable and loud for LISA at 1\,kpc, while DECIGO will be able to detect it
at Mpc scales.

Consistent with Figure~\ref{fig:A12-APN}, tidal distortion clearly dominates the PN contribution to the signal at the $n=1$ and 3 harmonics, 
providing a valuable constraint on the internal structure of the stars, especially if the system is close enough
to follow up with spectroscopic and/or photometric observations.
Note from Fig.~\ref{fig:A12-APN} that 
resonant forcing of the quadrupole $f$-mode is not significant.

\subsection{Signal from a $0.6\,M_\odot$ pair at 15\,mHz}
\label{section:example2}

\begin{figure}
  \centering
  \includegraphics[width=0.5\textwidth]{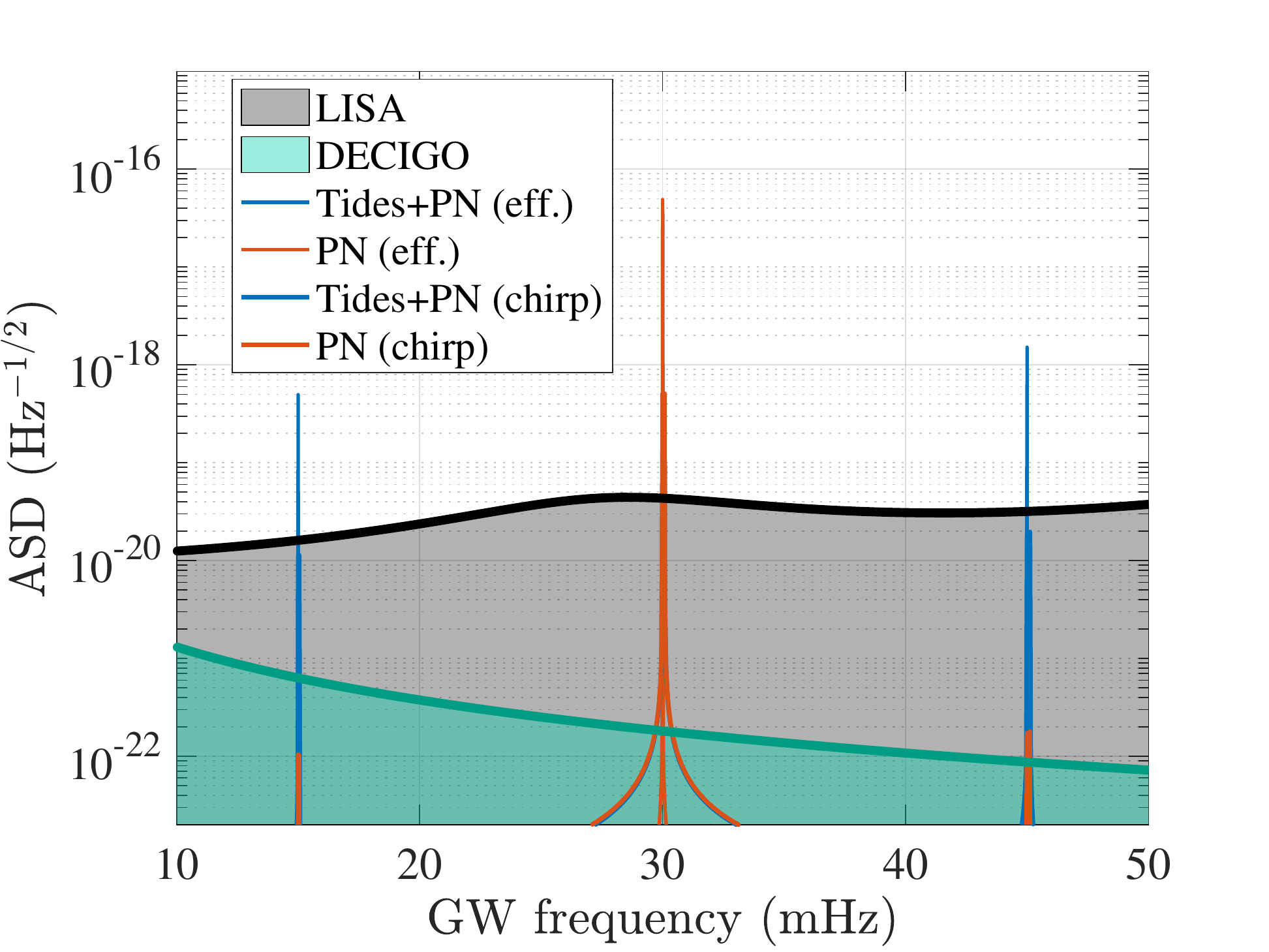}
  \caption{Same as Fig.~\ref{fig:30mHz} but for a quasi-circular pair of $0.6M_\odot$ white dwarfs with orbital frequency 15\,mHz, radii $10^9\,{\rm cm}$, again at a distance of 1 kpc and for 5 years of detector integration. 
Relative to the LISA noise ASD, the signal-to-noise
ratios are 80, 2300 and 130 for the $n=1,2,3$ harmonics respectively, again making the tidal contribution
to the signal resolvable and loud for LISA.
}
  \label{fig:15mHz}
\end{figure}

Figure~\ref{fig:15mHz} 
shows the true and effective ASD for a quasi-circular pair of $0.6\,M_\odot$ white dwarfs with orbital frequency 15\,mHz, radii $10^9\,{\rm cm}$, and again at a distance of $1\, \mathrm{kpc}$ and for 5 years of detector integration. 
Such a binary is close to filling its Roche lobe (see Fig.~\ref{fig:A12-APN}), and has a coalescence timescale of 1866 yr. 
While the chirp is small
at 0.1\,mHz for the $n=2$ harmonic, it significantly reduces the peak amplitudes of the three harmonics,
making it vital to use the chirped waveform \rn{eq:Hchirp} as a template in any optimal matched filter search. 
For this system, ${\cal S}/{\cal N}=80$, 2300 and 130 for the $n=1,2,3$ harmonics respectively for LISA, making the source detectable
at distances up to several kpc, while for DECIGO the corresponding values are 2000, $5 \times 10^5$ and $4\times10^4$, again making the system detectable at Mpc scales.

For this generic pair of white dwarfs, the tides again clearly dominate the PN contribution to the signal at the
$n=1$ and 3 harmonics. Note again from Fig.~\ref{fig:A12-APN} that 
resonant forcing of the quadrupole $f$-mode is not significant.

\subsection{Signal from the cataclysmic variable HM\,Cnc at 3\,mHz}\label{CV}

HM Cnc (RX J0806.3+1527) is a cataclysmic variable with an orbital period of 5.4 min (3.1\,mHz), making it
the shortest-period binary star known \citep{Barros2007} and therefore
one of the strongest ``verification sources'' of gravitational waves for LISA \citep{Roelofs2010}.
It consists of a white dwarf accreting from a less massive second white dwarf, with relative radial velocity measurements
providing a mass ratio estimate of $0.50\pm 0.13$ \citep{Roelofs2010}, 
and with upper limits on the masses of the individual components of 
$0.9\,M_\odot$ and $0.45\,M_\odot$ \citep{Barros2007}.
\citet{Roelofs2010} finds that the orbital frequency must be adjusted by $3.57\times 10^{-16}\,{\rm Hz\,s}^{-1}$
in order to improve the accuracy of the ephemeris over three years; this corresponds to masses of $0.54\,M_\odot$ and $0.27\,M_\odot$
if the frequency is modulated by gravitational wave emission alone. We adopt these masses here.

\begin{figure}
  \centering
 \includegraphics[width=0.5\textwidth]{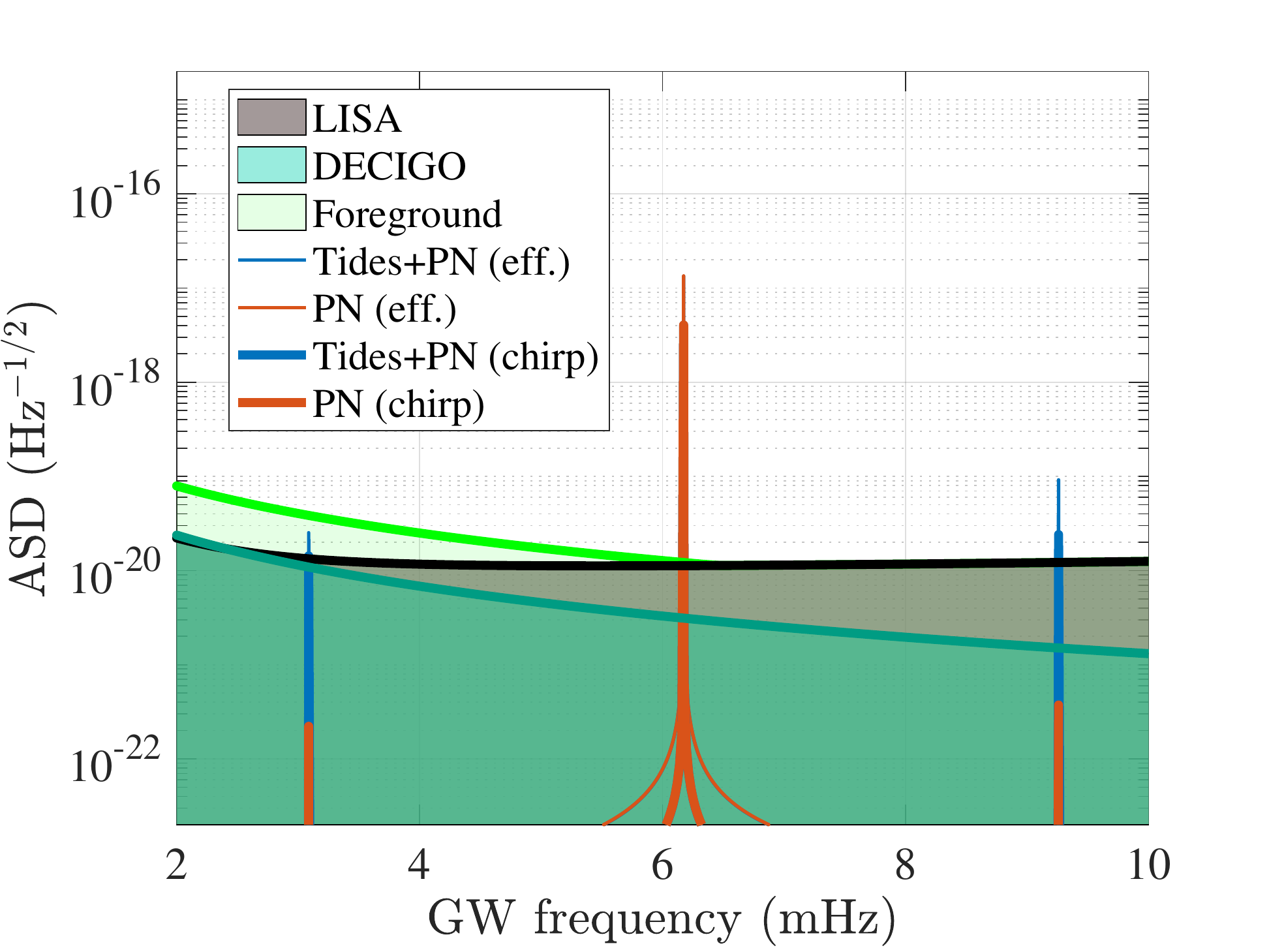}
  \caption{Effective ASD after 5 years of integration
  for the verification system HM Cnc, a 3.1\,mHz cataclysmic variable consisting of a white dwarf accreting
  from a companion white dwarf half its mass. Assuming a quasi-circular orbit, parameters adopted are $m_1=0.27\,M_\odot$, $m_2=0.54\,M_\odot$, 
  $R_1=2.1\times 10^9$\,cm (the Roche radius for this observationally determined mass ratio and period and adopted masses), and $D=1$\,kpc.
  Also shown is the foreground confusion limit adapted from \citet{Ruiter2010} Fig.~10 (light green curve), which drops below the LISA noise floor at around 6\,mHz. With a signal-to-noise of 20, the $n=3$ harmonic should be detectable at this distance.
}
  \label{fig:HMCnc}
\end{figure}

Excitation of the $n=1$ and 3 harmonics of the orbital period will persist for semi-detached and contact systems.
Figure~\ref{fig:HMCnc} 
shows the expected signal after 5 years of integration for donor and accretor masses of 
$0.27\,M_\odot$ and $0.54\,M_\odot$ respectively, donor radius of $R_1=2.1\times 10^9$\,cm (the Roche radius for this observationally determined mass ratio and period, assuming the adopted masses), and distance 1\,kpc. A quasi-circular orbit is assumed.
Also shown is the foreground confusion limit (light green curve) adapted from \citet{Ruiter2010} Fig.~10, showing that the foreground no longer dominates the noise around 6\,mHz (black curve).
With a signal-to-noise of 20, the $n=3$ harmonic should be detectable at this distance,
as it should be up to a distance of 2.3\,kpc for a signal-detection limit of 8. Note that over a 5 year observing period, the chirp is 56\,nHz from GR alone, which is greater than the frequency resolution $\Delta f$. The true signal therefore has finite width, and the resulting reduction in amplitude (compared to the effective ASD) is shown in the Figure.

Note that the chirp will be modified by mass transfer, which 
may well be non-conservative.
In fact the X-ray luminosity of this system is significantly lower than expected for a system in which mass-transfer is 
driven by gravitational radiation, and a possible way to reconcile this is if the system is losing angular momentum
through mass loss \citep{Willems2005}.

Note also that HM Cnc displays X-ray and optical light-curve modulations at the orbital period which are consistent with eccentricity 
of the order of 0.1 (Mardling, in preparation); if this is the case and the $n=3$ harmonic is detectable, its amplitude relative to that of the $n=2$
harmonic will be larger than shown here, and given the observational constraints on the masses, can be used to constrain the eccentricity according to our analysis (equation~\rn{zt0} with $e(0)> 0$). We will generalize our analysis to arbitrary eccentricity in a future publication.

Finally, note that due to the relatively large radius of the Roche-filling star, the tidal contribution to the gravitational wave signal from the $n=1$ and 3 harmonics is more than 100 times that of the general relativistic contribution.

\section{Summary and discussion}

Using the formalism of \citet{GM1980} for the self-consistent treatment of the interaction between a binary orbit and the dynamical tide of a non-rotating star, we have demonstrated that 
as long as the stars oscillate, the osculating eccentricity will be non-zero and the orbit will be non-circular as we show in equation~\rn{rt} and 
Figure~\ref{fig:eccev}.
Here we have only included the dominant quadrupole $f$-mode, neglecting any resonances that may exist with lower-frequency $g$-modes.

A small time-varying eccentricity produces gravitational wave power at the first and third harmonics of the binary frequency, in addition to the usual power in the second harmonic. The result is important because it provides access to knowledge of the internal structure of the stars, with the amplitude of the $n=1$ and 3 harmonics being proportional to the induced eccentricity, which itself depends linearly on the apsidal motion constants of the stars. Thus, in contrast to the classical technique used to determine these quantities \citep[eg.][]{Claret2010}, our approach does not necessitate the measurement of the rate of apsidal advance, and complements asteroseismological studies of pulsating white dwarfs \citep{Winget2008}. See also \citet{Seto2001b} and \citet{Willems2008} for the effect of apsidal advance on gravitational wave signals from binaries, and \citet{Batygin2009} and \citet{Mardling2010} for a description of how the apsidal motion constant of a hot Jupiter can be determined without measuring the apsidal motion rate.

Close binaries are expected to tidally circularize on a timescale which depends on the mass ratio, the ratio of the stellar radius to the stellar separation, the apsidal motion constants of the stars (which themselves depend on their equations of state), and their $Q$-values which characterize the efficiency at which they dissipate the tidal motion. We have assumed that the short-period systems considered here are in the relaxed quasi-circular state, with the corresponding gravitational wave signal depending on this assumption. However, it is not known how mass loss affects the orbital evolution, and there is reason to believe that semi-detached systems such as the LISA verification binary HM Cnc are not fully circularized (Mardling, in preparation). If this is indeed the case, it should be evident in the ratio of amplitudes of the $n=2$ and 3 harmonics, and in the frequency-splitting effect of the associated apsidal advance as described in \citet{Willems2008}. 

Another consequence of tidal evolution is spin synchronization and alignment. While our analysis does not include the influence of spin on the gravitational wave signal, one effect will be to contribute an additional quadrupole potential due to spin distortion, modifying
slightly the harmonic amplitudes.
Note that systems for which the spin synchronization timescale is longer than the orbital shrinkage timescale are not expected to actually achieve the synchronous state \citep{Fuller2012}, although it is likely to be close unless some kind of non-1:1 Cassini state is established which may happen if the eccentricity is significant and the equation of state is appropriate as is the case for Mercury in the Solar System
\citep[eg.][]{Correia2015}.

We have shown that the gravitational wave amplitudes for the $n=1$ and 3 harmonics are dominated by the tidal contribution to the potential, with the relativistic contribution at least an order of magnitude less for systems for which the these harmonics are detectable. In addition we have shown that the orbital frequencies of white dwarf binaries remain far from resonance with the dominant $f$-mode, all the way to Roche contact. As such, a measurement of the ratio of amplitudes of the $n=2$ and 3 harmonics corresponds to a measurement of the combination of parameters given in \rn{eq:both}, thereby placing useful constraints on the apsidal motion constants of the stars, especially if the masses and radii are available from standard spectroscopy and photometry as is the case for HM Cnc.

In spite of the fact that significantly non-circular short-period white dwarf binaries are expected to be extremely rare in the Galaxy,
significant effort has been spent on calculating their gravitational wave signal \citep[eg.][]{Pierro2001,Willems2007}. While the dense stellar environments in which such binaries form promote the formation of highly eccentric binaries, and indeed excite the eccentricities of tidally relaxed binaries through multiple encounters, in the end it is a contest between the circularization timescale and the encounter timescale, with the former almost certainly winning in most cases. On the other hand, while the vast majority of close binary white dwarfs are isolated in the field and are therefore expected to be circular, the work described here offers the chance to test this assumption using the new frontier of gravitational wave astronomy.

\section*{Acknowledgements}
LM acknowledges support by an Australian Government Research Training (RTP) Scholarship. BM has been supported, in part, by the Australian Research Council through a Future Fellowship (FT160100035). Mathematica was used to perform parts of the analysis.

%%%%%%%%%%%%%%%%%%%%%%%%%%%%%%%%%%%%%%%%%%%%%%%%%%



%\bibliographystyle{mnras}

% This method is tedious and prone to error if you have lots 
\bibliography{refs}{99}       %use a bibtex bibliography file refs.bib
\bibliographystyle{mnras}
%\bibliographystyle{apalike} 
%%%%%%%%%%%%%%%%%%%%%%%%%%%%%%%%%%%%%%%%%%%%%%%%%%

%%%%%%%%%%%%%%%%% APPENDICES %%%%%%%%%%%%%%%%%%%%%

\appendix

\section{The rate of change of $\lowercase{z}=\lowercase{e} {\rm \lowercase{e}}^{\lowercase{i}\varpi}$}
\label{appendix:zdot}
In the case that the perturbing acceleration can be written in terms of the gradient of a potential, one can use
Lagrange's planetary equations \citep[eg.][]{murray} to write
\bea
\dot z
&=&\left(\frac{\dot e}{e}+i\dot\varpi\right)z 
=\frac{1}{\mu\nu a^2 e}\left(-\frac{1}{e}\pp{\cal R}{\varpi}
+i\pp{\cal R}{e}\right)z.
\eea
Using the chain rule we have
\bea
\pp{\cal R}{e}&=&\pp{\cal R}{z}\pp{z}{e}+\pp{\cal R}{z^*}\pp{z^*}{e}
=\frac{1}{e}\left(z\pp{\cal R}{z}+z^*\pp{\cal R}{z^*}\right)
\eea
and
\bea
\pp{\cal R}{\varpi}&=&\pp{\cal R}{z}\pp{z}{\varpi}+\pp{\cal R}{z^*}\pp{z^*}{\varpi}
=i\left(z\pp{\cal R}{z}-z^*\pp{\cal R}{z^*}\right)
\eea
so that
\be
\dot z=\frac{2i}{\mu\nu a^2}\pp{{\cal R}}{z^*}.
\ee
The rate of apsidal advance is then
\be
\dot\varpi=\frac{1}{2i}\left(\frac{\dot z}{z}-\frac{\dot z^*}{z^*}\right).
\label{wdotz}
\ee
For a general perturbing acceleration ${\bf a}$, the rate of change of $z$ can be written in terms of the Runge-Lenz vector, the latter being \citep[eg.][]{RM2002}
\be
\dot{\bf e}=\left(2({\bf a}\cdot\dot{\bf r}){\bf r}-({\bf r}\cdot\dot{\bf r}){\bf a}-({\bf a}\cdot{\bf r})\dot{\bf r}\right)/G(m_1+m_2),
\ee
so that using equations (28) and (31) of \citet{RM2002} with
$\hat{\bf e}=\cos\varpi\,{\bf i}+\sin\varpi\,{\bf j}$ and $\hat{\bf q}=-\sin\varpi\,{\bf i}+\cos\varpi\,{\bf j}$
we have

\be
\dot z=({\bf i}\cdot\dot{\bf e})+i\,({\bf j}\cdot\dot{\bf e}).
\ee
In terms of the mean anomaly $M$,
to first-order in eccentricity one has
\be
r=a(1-e\cos M), \hspace{0.5cm} \dot r=a\nu e\sin M,
\ee
and
\be
f=M+2e\sin M, \hspace{0.5cm} \dot f=\nu(1+2e\cos M).
\ee
Putting ${\bf e}_r=\cos(f+\varpi){\bf i}+\sin(f+\varpi){\bf j}$,
${\bf e}_\psi=-\sin(f+\varpi){\bf i}+\cos(f+\varpi){\bf j}$ and $M=\lambda-\varpi$, for ${\bf a}={\bf a}_{\rm PN}$ one obtains

\be
\dot z_{\rm PN}=i\nu\left(\frac{a\nu}{c}\right)^2\left[3z-(3-\eta){\rm e}^{i\lambda}-(11-6\eta)z^*{\rm e}^{2i\lambda}\right],
\label{wGR}
\ee
while for ${\bf a}=(1/\mu)\partial{\cal R}_{\rm tide}/\partial{\bf r}$ (see equation~\rn{rdd}) we have

\be
\dot z_{\rm QD}=i\nu k_2\left(\frac{m_2}{m_1}\right)\left(\frac{R_1}{a}\right)^5
\left[15z+6 {\rm e}^{i\lambda}+27z^* {\rm e}^{2i\lambda}\right]
\label{wtide}
\ee
which recover \rn{zdot} to zeroth-order in $z$ and $z^*$ (in the limit that $\nu\ll\omega_{12}$). On the other hand, orbit-averaging gives
\be
\left<\dot z\right>=\left<\dot z_{\rm QD}+\dot z_{\rm PN}\right>
=i(\dot\varpi^{\rm (QD)}_{\rm sec}+\dot\varpi^{\rm (PN)}_{\rm sec})z,
\label{zdotsec}
\ee
where
\be
\dot\varpi^{\rm (QD)}_{\rm sec}=3\nu\left(\frac{a\nu}{c}\right)^2
\hspace{0.3cm}{\rm and}\hspace{0.3cm}
\dot\varpi^{\rm (PN)}_{\rm sec}=15\nu k_2\left(\frac{m_2}{m_1}\right)\left(\frac{R_1}{a}\right)^5
\ee
are the standard expressions for the secular rate of apsidal advance due to 
the tidal quadrupole distortion and post-Newtonian acceleration \citep[eg.][]{Sterne1939,RM2002}, so that from \rn{wdotz},
\be
\dot\varpi_{\rm sec}=\dot\varpi^{\rm (QD)}_{\rm sec}+\dot\varpi^{\rm (PN)}_{\rm sec}.
\label{wsec}
\ee
Note that since integrating \rn{zdotsec} gives
\be
z(t)=z(0){\rm e}^{i\dot\varpi_{\rm sec} t},
\ee
there is no secular variation associated with quasi-circular orbits (for which $z(0)=0$).

%\newpage

%\section{Notation}

\newpage
\begin{table}
\caption{Notation}
	\label{notation}
	\begin{tabular}{lll}
		 Symbol & Definition & Reference/definition \\ \hline
		$A_{12}$ & Tidal contribution to eccentricity amplitude & Eq. \rn{eq:A12} \\
		$A_{\mathrm{PN}}$ & Post-Newtonian contribution to eccentricity amplitude  & Eq. \rn{eq:APN} \\
		$a$ & Semi-major axis & \\
		$b_{\textbf{k}}(t)$ & Oscillation amplitude & Eqs. \rn{btw} and \rn{bt} \\
		$c_{l\!m}$ & Spherical harmonic constants & $c_{20}=-\sqrt{\pi/5}$, $c_{22}=c_{2-2}=\sqrt{3\pi/10}$ \\
		$D$ & Distance to binary  &  \\
		$e$ & eccentricity &  \\
		$\textbf{e}$ & Runge-Lenz vector & \cite{goldstein} \\
		$E_{\mathrm{tot}}$ & Total energy of coupled orbit and oscillator & \cite{RM1995a,RM1995b,Lai1997} \\
		$E_{\mathrm{PN}}$ & Post-Newtonian energy & \cite{kidder1995}  \\
		$f$ & True anomaly &  \\
		$H$ & Continuous fourier transform of signal &  \\
		$h_{i\!j}$ & Dimensionless gravitational wave strain tensor & \cite{Einstein1918,Kokkotas1999} \\
		$h_0$ & Dimensionless strain amplitude from binary at distance $D$ & $h_0 = 2\eta\left(a/D\right)\left(a\nu/c\right)^4$  \\
		$h_+$ & Strain in plus polarisation & Eq. \rn{eq:hplus} \\
		$h_\times$ & Strain in cross polarisation  & Eq. \rn{eq:hcross} \\
		$\tilde{h}$ &  Strain amplitude spectral density (Hz$^{-1/2}$)& Eq. \rn{eq:htilde} \\
		$I_{k\!l}$ & Structure constant, dimensionless moment of inertia of eigenmode &  \cite{GM1980,RM1995a} \\
		$\mathcal{I}^{i\!j}$ & Quadrupole moment tensor & \cite{Blanchet1990} \\
		$\textbf{J}_{\mathrm{tot}}$ & Total angular momentum of coupled orbit and oscillator & \cite{RM1995a,RM1995b} \\
		$\textbf{J}_{\mathrm{PN}}$ &  Post-Newtonian angular momentum& \cite{kidder1995} \\
		$k_2$ & Apsidal motion constant &  \cite{Sterne1939} \\
		$k$ &  Mode radial order &  \\
		$l$ &  Mode spherical harmonic degree &  \\
		$m$ & Mode spherical harmonic order &  \\
	    	$\textbf{k}$ & Mode number & $klm$ \\
		$m_i$ & Mass of $i$th star  &  \\
		$\mathcal{M}$ & Chirp mass of binary & $(m_1m_2)^{3/5}/(m_1+m_2)^{1/5}=\eta^{3/5} (m_1+m_2)$\\
		$M$ & Mean anomaly &  \\
		$Q_i$ & Tidal quality factor ($Q$-value) for $i$th star & \cite{Piro2011} \\
		$Q_{k\!l}$ & Overlap integral & \cite{Lee1986} \\
		$r$ & Stellar separation  &  \\
		$\textbf{r}$ & Position vector of $m_2$ relative to $m_1$ &  \\
		$R_{i}$ &  Radius of star $i$ &  \\
		$\mathcal{R}_{\mathrm{tide}}$ & Tide--orbit interaction energy & \cite{RM1995a} \\		$\mathcal{S}/\mathcal{N}$ & Matched filer signal--to--noise ratio  & \cite{Flanagan1998,Moore2015} \\
		$S_f$ & Instrumental detector noise (Hz$^{-1}$) &  \\
		$T_{k\!l}$ & Structure constant & \cite{RM1995a,GM1980} \\
		$T$ & Observing time &  \\
		$X_n^{-(l+1),m}(e)$ & Hansen coefficient & Eq. \ref{hansen}, \cite{murray} \\
		$x_n^{-(l+1),m}$ & Coefficient of leading order term of Hansen coefficient   &Table \ref{xn}, \cite{Mardling2013} \\
		$x^i$ & Components of $\textbf{r}$ &  \\
		$z$ & Complex eccentricity & $z=e{\rm e}^{i\varpi}$, \cite{Laskar12} \\
		$\eta$ & Dimensionless symmetric mass ratio  & $\eta = m_1m_2/(m_1+m_2)^2$ \\
		$\lambda$ & Mean longitude & $\lambda = M + \varpi$ \\
		$\Lambda_{i\!j,k\!l}$ & Sky--projection operator & \cite{Maggiore2007} \\
		$\mu$ & Reduced mass & $\mu = m_1 m_2/(m_1+m_2)$ \\
		$\nu$ &  Orbital frequency (rad s$^{-1}$)&  \\
		$\overline\nu$ & Dimensionless orbital frequency  & $\nu/(G(m_1+m_2)/R_1)^{1/2}$\\
		$\varpi$ & Longitude of periastron  &  \\
		$\tau_{\mathrm{tides}}$ & Tidal damping timescale & \cite{Hut1981}  \\
		$\tau_{\mathrm{GR}}$ & Merger timescale from gravitational wave emission & \cite{Peters1964}\\
		$\tau_\nu$ & Chirp timescale & Eq. \ref{eq:tauchirp} \\
		$\Phi_{\mathrm{quad}}$ & Quadrupolar perturbing potential & \cite{Sterne1939} \\
		$\psi$ & True longitude & $\psi= f + \varpi$  \\
		$\omega_{k\!l}$ & Oscillation frequency of mode with triplet $\textbf{k}$ & \\
		$\overline{\omega}_{k\!l}$ & Dimensionless oscillation frequency & ${\omega}_{k\!l}/(G (m_1+m_2)/R_1^3)^{1/2}$   \\
	\end{tabular}
	\end{table}

% Don't change these lines
\bsp	% typesetting comment
\label{lastpage}
\end{document}